\documentclass{svmult}

\usepackage{amsmath}
\usepackage{amssymb}
\usepackage{color}

\usepackage{mathptmx}       
\usepackage{helvet}         
\usepackage{courier}        
\usepackage{type1cm}        
\usepackage{makeidx}         
\usepackage{graphicx}        
\usepackage{multicol}        
\usepackage[bottom]{footmisc}

\makeindex             

\usepackage{amsmath,amssymb,mathrsfs, algorithm2e}

\usepackage{subfig,graphicx}

\newcommand{\Pochhammer}[2]{{\left( #1\right) }_{#2}}

\newcommand{\dfield}[2]{(#1,#2)}
\newcommand{\const}[2]{\text{const}_{#2}#1}

\newcommand{\lr}[1]{\langle#1\rangle}
\newcommand{\ltr}[1]{[#1,#1^{-1}]}

\newcommand{\depth}{\delta}

\newcommand{\sigmaE}{$\Sigma$}
\newcommand{\pisiE}{$\Pi\Sigma$}
\newcommand{\rpisiE}{$R\Pi\Sigma$}
\newcommand{\rpiE}{$R\Pi$}
\newcommand{\piE}{$\Pi$}
\newcommand{\rE}{$R$}

\newcommand{\BRSEE}{\mathbb{E}}
\newcommand{\NN}{\mathbb{N}}
\newcommand{\ZZ}{\mathbb{Z}}
\newcommand{\AR}{\mathbb{A}}
\newcommand{\KK}{\mathbb{K}}
\newcommand{\QQ}{\mathbb{Q}}
\newcommand{\GG}{\mathbb{G}}

\newcommand{\HH}{\mathbb{H}}
\newcommand{\FF}{\mathbb{F}}
\newcommand{\vect}[1]{\mathbf{#1}}

\newcounter{linectr}
\newcommand{\algname}[1]{{\tt #1}}
\newenvironment{alg}[5]
{\begin{Algorithm}#1\makebox[0.0cm][l]{}\small{#2}

\vspace*{0.0cm}\noindent \algname{#3}\\
\vspace*{0.1cm}\noindent\hspace*{0.1cm}\makebox[1.1cm][l]{\sf
Input:}\begin{minipage}[t]{10.5cm}{#4}\end{minipage}\\
\vspace*{0.1cm}\noindent\hspace*{0.1cm}\makebox[1.1cm][l]{\sf
Output:}\begin{minipage}[t]{10.5cm}{#5}\end{minipage}\\
\vspace*{0.cm}
\begin{list}{(\arabic{linectr})}{\usecounter{linectr} \labelwidth3ex\itemsep0ex\labelsep1ex\leftmargin5ex\parskip-0.2cm
\listparindent0ex}} {\end{list}\end{Algorithm}\normalsize}

\newtheorem{Algorithm}{Algorithm}

\newcounter{mmacnt}
\def\restartmma{\setcounter{mmacnt}{0}}
\restartmma \catcode`|=\active
\def|#1|{\mathrm{#1}}
\catcode`|=12
\newenvironment{mma}{
 \par\smallskip
 \catcode`|=\active
 \parskip=0pt\parindent=0pt 
 \small
 \def\In##1\\{%
   \def\linebreak{\hfill\break\null\qquad}%
   \refstepcounter{mmacnt}
   \hangindent=2.5em\hangafter=0
   \leavevmode
   \llap{\tiny\sffamily In[\arabic{mmacnt}]:=\kern.5em}%
   \mathversion{bold}\footnotesize$\tt\bf\displaystyle##1$\normalsize
   \mathversion{normal}\par
 }%
 \def\Print##1\\{%
   \def\linebreak{\hfill\break}%
   \hangindent=2.5em\hangafter=0
   \leavevmode\footnotesize ##1\par}%
 \def\Out##1\\{%
   \def\linebreak{$\hfill\break\null\hfill$}%
   \kern\abovedisplayskip\par
   \hangindent=2.5em\hangafter=0
   \leavevmode
   \llap{\tiny\sffamily Out[\arabic{mmacnt}]=\kern.5em}
   \footnotesize$\displaystyle\tt##1$\normalsize\hfill\null\par
   \kern\belowdisplayskip
 }%
 \def\Warning##1##2\\{%
   \def\linebreak{\hfill\break}%
   \hangindent=2.5em\hangafter=0
   \leavevmode
   {\scriptsize##1 : ##2}\par}%
}{%
 \par\smallskip
}

\spnewtheorem{mtheorem}{Theorem}[section]{\bfseries}{\itshape}
\spnewtheorem{mdefinition}{Definition}[section]{\bfseries}{}
\spnewtheorem{mremark}{Remark}[section]{\bfseries}{}
\spnewtheorem{mlemma}{Lemma}[section]{\bfseries}{\itshape}
\spnewtheorem{mexample}{Example}[section]{\bfseries}{}
\spnewtheorem{mcorollary}{Corollary}[section]{\bfseries}{\itshape}
\spnewtheorem{mproposition}{Proposition}[section]{\bfseries}{\itshape}

\begin{document}

\title*{Refined Holonomic Summation Algorithms in Particle Physics
\bigskip
{\small \text{Dedicated to Sergei A. Abramov on the occasion of his 70th birthday}}}
 \titlerunning{Refined Holonomic Summation Algorithms in Particle Physics}
\author{Johannes Bl\"umlein, Mark Round and Carsten Schneider}
\institute{Johannes Bl\"umlein \at Deutsches Elektronen-Synchroton (DESY), Zeuthen, Platanenalle 6, D-15738 Zeuthen, Germany, \email{Johannes.Bluemlein@desy.de}
\and Mark Round \and Carsten Schneider \at Research Institute for Symbolic Computation (RISC), Johannes Kepler University Linz, Altenberger Stra\ss e 69, A-4040, Linz, Austria, \email{mround@risc.jku.at,cschneid@risc.jku.at}
}
%
%
\maketitle

\abstract{An improved multi-summation approach is introduced and discussed that enables one to simultaneously handle sequences generated by indefinite nested sums and products in the setting of difference rings and holonomic sequences described by linear recurrence systems.
Relevant mathematics is reviewed and the underlying advanced difference ring machinery is elaborated upon. The flexibility of this new toolbox contributed substantially to evaluating complicated multi-sums coming from particle physics.  Illustrative examples of the functionality of the new software package~\texttt{RhoSum} are given.
}

\section{Introduction\label{sec:intro}}

A standard approach to symbolic summation is that of telescoping.  There are many individual variations and specific technologies, however one may summarize all technologies by stating the problem of parameterized telescoping.  Given sequences $F_1(k),\dots,F_d(k)$ over an appropriate field $\KK$, find $d$ constants (meaning free of $k$ and not all zero) $c_1,\dots,c_d\in\KK$ and a sequence $G(k)$ such that
\begin{equation}\label{eq:ptcert}
G(k+1)-G(k)  = c_1 F_1(k) + \ldots + c_d F_d(k).
\end{equation}
If one does succeed then one can sum the relation to obtain
\begin{equation}\label{eq:ptproblem}
G(b+1) - G(a)=c_1 \sum_{k=a}^b F_1(k) + \ldots + c_d \sum_{k=a}^b F_d(k)  .
\end{equation}
for some properly chosen bounds $a,b$.  Restricting to $d=1$ gives the telescoping formula, which expresses the sum over $F_1$ as a difference.  Alternatively, suppose that we started with the definite sum $S(n)=\sum_{k=l}^{L(n)}F(n,k)$ with a bivariate sequence $F(n,k)$, with $l\in\NN$ and with some integer linear expression\footnote{$L(n_1,\dots,n_l)$ stands for $z_0+z_1\,n_1+\dots+z_l\,n_l$ for some integers $z_0,\dots,z_l$.} $L(n)$. Then taking $F_i(k) := F(n+i-1,k)$ for $1\leq i\leq d$, the parameterized telescoping equation~\eqref{eq:ptproblem} reduces to Zeilberger's creative telescoping. Namely, omitting some mild assumptions, equation~\eqref{eq:ptproblem} with $a=l$ and $b=L(n)$ yields a linear recurrence of the form
\begin{equation}\label{Equ:Rec}
h(n)=c_1(n)\, S(n)+c_2(n)\, S(n+1)+\dots+c_d(n)\, S(n+d-1)
\end{equation}
where $h(n)$ comes from $G(L(n)+1) - G(l)$ and extra terms taking care of the shifts in the boundaries. In the creative telescoping setting we assume that the field $\KK$ contains the variable $n$ and thus the constants $c_i$ may depend on $n$.

This recurrence finding technology based on parameterized telescoping started for hypergeometric sums~\cite{Gosper1978,Zeilberger, petkovsek1996b,PS:95} and has been extended to $q$-hypergeometric and their mixed versions~\cite{BAUER1999711, PR:97}. A generalization to multi-summation has also been performed~\cite{WZtheory,Wegschaider,AZ:06}. Further, the input class has been widened significantly by the holonomic summation paradigm~\cite{Zeilberger:90a} and the efficient algorithms worked out in~\cite{CHYZAK2000115,Koutschan:10}. Using these tools it is possible to solve the parameterized telescoping problem for (multivariate) sequences that are a solution of a system of linear difference (or differential) equations. In particular, applying this technology recursively enables one to treat  multi-sum problems.

Another general approach was initiated by M.~Karr's summation algorithm~\cite{Karr:81}
in the setting of \pisiE-fields and has been generalized to the more general setting of \rpisiE-difference ring extensions~\cite{DR1,DR2,DR3}. Using the summation package \texttt{Sigma}~\cite{Schneider:07} one can solve the parameterized telescoping equation in such difference rings and fields not only for the class of \hbox{($q$--)}hypergeometric products and their mixed versions, but also for indefinite nested sums defined over such objects covering as special cases, e.g., the generalized harmonic sums~\cite{Ablinger:2013cf}
\begin{equation}
S_{r_1,\ldots,r_m}(x_1,\dots,x_m,n) = \sum_{k_1=1}^n\frac{x^{k_1}}{k_1^{r_1}}\sum_{k_2=1}^{k_1}\frac{x^{k_2}}{k_2^{r_2}}\cdots\sum_{k_m=1}^{k_{m-1}}\frac{x^{k_m}}{k_m^{r_m}}
\end{equation}
with $x_1,\dots,x_m\in\KK\setminus\{0\}$ which contain as special case the so-called harmonic
sums~\cite{Blumlein:1998if,Vermaseren:99} defined by
$S_{x_1\,r_1,\ldots,x_m\,r_m}(n)=S_{r_1,\ldots,r_m}(x_1,\dots,x_m,n)$ with $x_1,\dots,x_r\in\{-1,1\}$.
Further, \rpisiE-extensions enable one to model cyclotomic sums~\cite{ABS:11} or nested binomial sums~\cite{ABRS:14}. Using efficient recurrence solvers~\cite{Petkov:92,Bron:00,Schneider:01,Schneider:05b,ABPS:17} that make use of d'Alembertian solutions~\cite{Abramov:94,Abramov:96}, a strong machinery has been developed to transform definite sums to expressions in terms of indefinite nested sums defined over \hbox{($q$--)}hypergeometric products and their mixed versions.  (d'Alembertian solutions are a subclass of Liouvillian solutions~\cite{Singer:99,Petkov:2013}.) In the last years this strong machinery~\cite{Schneider:13} has been utilized heavily for problems in particle physics, see, e.g.,~\cite{Physics1,Physics2,Physics3,Physics4} and references therein. However, for several  instances we were forced to push forward our existing summation technologies to be able to carry out our calculations.

More precisely, we utilized and refined the Sigma-approach that has been developed in~\cite{Schneider:05a,Schneider:07} to unite Karr's \pisiE-field setting with the holonomic system approach: one can solve the parameterized telescoping problem in terms of elements from a \pisiE-field together with summation objects which are solutions of inhomogeneous linear difference equations. In particular, a refined tactic has been worked out for the well-known holonomic approach~\cite{CHYZAK2000115} that finds recurrences without Gr\"obner basis computations or expensive uncoupling algorithms~\cite{Zuercher:94,BCP13}. This efficient and flexible approach has been applied to derive the first alternative proof~\cite{APS:05} of Stembridge's TSPP theorem~\cite{STEMBRIDGE1995227}.

This article is the continuation of this work and explains new features that were necessary to compute highly non-trivial problems coming from particle physics~\cite{Physics1,Physics2,Physics3,Physics4}. First, the ideas of~\cite{Schneider:05a} are generalized from the difference field to the ring setting: we consider a rather general class of difference rings that is built by the so-called \rpisiE-extensions~\cite{DR1,DR2} and introduce on top a so-called higher-order linear difference extension. In this way, indefinite nested sums can be defined covering in addition a summation object that is a solution of an (inhomogeneous) linear difference equation defined over indefinite nested sums and products. In particular, our new techniques from~\cite{Schneider:08,Schneider:14,DR1} are applied to derive new and more flexible algorithms for the parameterized telescoping problem. Further, we push forward the theory of higher-order linear extensions in connection with \rpisiE-extension. We show that certain non-trivial constants, in case of existence, can be computed in such rings and that such constants can be utilized to design improved higher-order linear extension with smaller recurrence order. Finally, this machinery is applied recursively to multi-sums in order to produce linear recurrences.
As it turns out, our refined difference ring algorithms in combination with the ideas from~\cite{Schneider:05a}
introduce various new options as to how such recurrences can be calculated: using different telescoping strategies will lead to more or less complicated recurrence relations and the calculation time might vary heavily. In order to dispense the user from all these considerations, a new summation package~\texttt{RhoSum} has been developed that analyzes the different possibilities by clever heuristics and performs the (hopefully) optimal calculation automatically.

The manuscript is organized as follows.  In Section~\ref{DFTheory} we present our toolbox to solve the parameterized telescoping problem in our general setting built by \pisiE-fields, \rpisiE-extensions and a higher-order linear extension on top. In addition, new theoretical insight is provided that connects non-trivial constants in such extensions and the possibility to reduce the recurrence order of higher-order extensions.
Then, in Section~\ref{Sec:MultiSumApproach}, our multi-sum approach based on our refined holonomic techniques is presented and and specific technical aspects of the algorithm are explained.  Such details are important for an efficient implementation. With the main results of the summation approach discussed, Section~\ref{Examples} gives an illustrative example arising from particle physics which shows some of the features of the algorithm in practice. Finally, in Section~\ref{Summ} there is a brief summary.

\section{\label{DFTheory} Parameterized telescoping algorithms in difference rings}\label{Sec:DRTools}

In our difference ring approach the summation objects are represented by elements in a ring or field $\AR$
and the shift operator acting on these objects is rephrased in terms of a ring or field automorphism $\sigma:\AR\to\AR$.
In short, we call $\dfield{\AR}{\sigma}$ a difference ring or a difference field. The set of units of a ring $\AR$ is denoted by $\AR^*$ and the set of constants of $\dfield{\AR}{\sigma}$ is defined by
$$\const{\AR}{\sigma}=\{f\in\AR\mid\sigma(f)=f\}.$$
In general $\KK:=\const{\AR}{\sigma}$ is a subring of $\AR$ (or a subfield of $\AR$ if $\AR$ is field). In the following we will take care that $\KK$ is always a field containing the rational numbers $\QQ$ as subfield. $\KK$ will be also called the constant field of $\dfield{\AR}{\sigma}$. For a vector $\vect{f}=(f_1,\dots,f_d)\in\AR^d$ we define $\sigma(\vect{f})=(\sigma(f_1),\dots,\sigma(f_d))\in\AR^d$. $\NN$ denotes the non-negative integers.

Finally, we will heavily use the concept of difference ring
extensions. A difference ring $\dfield{\BRSEE}{\sigma'}$ is a
difference ring extension of a difference ring
$\dfield{\AR}{\sigma}$ if $\AR$ is a subring of $\BRSEE$ and
$\sigma'(f)=\sigma(f)$ for all $f\in\AR$. If it is clear from the
context, we will not distinguish anymore between $\sigma$ and
$\sigma'$.

Suppose that one succeeded in rephrasing the summation objects $F_1(k),\dots,F_d(k)$ in a difference ring $\dfield{\AR}{\sigma}$ with constant field $\KK$, i.e., $F_i(k)$ can be modeled by $f_i\in\AR$ where the corresponding objects $F_i(k+1)$ correspond to the elements $\sigma(f_i)$. Then the problem of parameterized telescoping~\eqref{eq:ptcert} can be rephrased as follows.

\begin{center}
 \noindent\fbox{\begin{minipage}{11.4cm}
\noindent \textbf{Problem RPT in $\dfield{\AR}{\sigma}$:} Refined Parameterized Telescoping.\\
\small
\textit{Given} a difference ring $\dfield{\AR}{\sigma}$ with constant field $\KK$ and $\vect{f}=(f_1,\dots,f_d)\in\AR^d$.\\
\textit{Find}, if possible, an\footnote{The optimality criterion
will be specified later in the setting of \rpisiE-extensions.}
``optimal'' difference ring extension $\dfield{\BRSEE}{\sigma}$ of
$\dfield{\AR}{\sigma}$ with
$\const{\BRSEE}{\sigma}=\const{\AR}{\sigma}$, $g\in\BRSEE$ and
$c_1,\dots,c_d\in\KK$ with $c_1\neq0$ and
\begin{equation}\label{Equ:ParaDF}
\sigma(g)-g=c_1\,f_1+\dots+c_d\,f_d.
\end{equation}
\end{minipage}}
\end{center}

\noindent Namely, suppose that we find  $g\in\BRSEE$ and
$c_1,\dots,c_d\in\KK$ and we succeed in reinterpreting $g$ as $G(k)$
in terms of our class of summation objects where $\sigma(g)$
represents $G(k+1)$. Then this will lead to a solution of the
parameterized telescoping equation~\eqref{eq:ptcert}. In
Subsection~\ref{Subsec:PureRPiSi} we will work out this machinery
for difference rings that are built by
\rpisiE-extensions~\cite{DR1,DR2,DR3} and which enables one to model
summation objects, e.g., in terms of indefinite nested sums defined
over \hbox{($q$--)}hypergeometric products and their mixed versions.
Afterwards, we will use this technology in
Subsection~\ref{Subsec:RPTHolo} to tackle summation objects that can
be represented in terms of recurrences whose coefficients are given
over the earlier defined difference rings.

\subsection{\pisiE-fields and \rpisiE-extensions}\label{Subsec:PureRPiSi}

A central building block of our approach are Karr's \pisiE-fields~\cite{Karr:81,Karr:85}.

\begin{mdefinition}
A difference
field $\dfield{\FF}{\sigma}$ with constant field $\KK$ is called a
\pisiE-field if $\FF=\KK(t_1,\dots,t_e)$ where for all $1\leq i\leq
e$ each $\FF_i=\KK(t_1,\dots,t_i)$ is a transcendental field
extension of $\FF_{i-1}=\KK(t_1,\dots,t_{i-1})$ (we set $\FF_0=\KK$)
and $\sigma$ has the property that
$\sigma(t_i)=a\,t_i$ or
$\sigma(t_i)=t_i+a$ for some $a\in\FF_{i-1}^*$.
\end{mdefinition}

\begin{mexample}\label{Exp:BaseFields}
(1) The simplest \pisiE-field is the rational difference field $\FF=\KK(t)$ for some field $\KK$ and the field automorphism $\sigma:\FF\to\FF$ defined by $\sigma(c)=c$ for all $c\in\KK$ and $\sigma(t)=t+1$. \\
(2) Another \pisiE-field is the $q$-rational difference field. Here one takes a rational function field $\KK=\KK'(q)$ over a field $\KK'$ and the rational function field $\FF=\KK(t)$ over $\KK$. Finally, one defines the field automorphism $\sigma:\FF\to\FF$ by $\sigma(c)=c$ for all $c\in\KK$ and $\sigma(t)=q\,t$.\\
(3) One can combine the two constructions (1) and (2) and arrives at the mixed $(q_1,\dots,q_e)$-multibasic rational difference field~\cite{BAUER1999711}. Here one considers the rational function field $\KK=\KK'(q_1,\dots,q_e)$ over the field $\KK'$ and the rational function field $\FF=\KK(t,t_1,\dots,t_e)$ over $\KK$. Finally, one takes the field automorphism $\sigma:\FF\to\FF$ determined by $\sigma(c)=c$ for all $c\in\KK$, $\sigma(t)=t+1$ and $\sigma(t_i)=q_i\,t_i$ for all $1\leq i\leq e$. By~\cite[Cor.~5.1]{OS:17} $\dfield{\FF}{\sigma}$ is again a \pisiE-field. \\
(4) Besides these base fields, one can model nested summation objects. E.g., one can can define the \pisiE-field $\dfield{\KK(t)(h)}{\sigma}$ with constant field $\KK$ where $\dfield{\KK(t)}{\sigma}$ is the rational difference field and $\sigma$ is extended from $\KK(t)$ to the rational function field $\KK(t)(h)$ subject to the relation $\sigma(h)=h+\frac1{t+1}$. Here $h$ in $\FF$ scopes the shift behavior of the harmonic numbers $S_1(k)=\sum_{i=1}^k\frac1i$ with $S_1({k+1})=S_1(k)+\frac1{k+1}$.
\end{mexample}

A drawback of Karr's very elegant \pisiE-field construction is the inability to treat the frequently used summation object $(-1)^k$. In order to overcome this situation, \rpisiE-extensions have been introduced~\cite{DR1,DR2,DR3}.

\begin{mdefinition}
A difference ring $(\BRSEE,\sigma)$ is called an
\emph{$APS$-extension} of a difference ring $\dfield{\AR}{\sigma}$
if $\AR=\AR_0\leq\AR_1\leq\dots\leq\AR_e=\BRSEE$ is a tower of ring
extensions where for all $1\leq i\leq e$ one of the following holds:
\begin{itemize}
 \item $\AR_i=\AR_{i-1}[t_i]$ is a ring extension subject to the relation $t_i^n=1$ for some $n>1$ where $\frac{\sigma(t_i)}{t_i}\in(\AR_{i-1})^*$ is a primitive  $n$th root of unity \emph{($t_i$ is called an $A$-monomial, and $n$ is called the order of the $A$-monomial)};
 \item $\AR_i=\AR_{i-1}[t_i,t_i^{-1}]$ is a Laurent polynomial ring extension with $\frac{\sigma(t_i)}{t_i}\in(\AR_{i-1})^*$ \emph{($t_i$ is called a $P$-monomial)};
 \item $\AR_i=\AR_{i-1}[t_i]$ is a polynomial ring extension with $\sigma(t_i)-t_i\in\AR_{i-1}$ \emph{($t_i$ is called an $S$-monomial)}.
\end{itemize}
If all $t_i$ are $A$-monomials, $P$-monomials or $S$-monomials, we
call $\dfield{\BRSEE}{\sigma}$ also a (nested) $A$-extension,
$P$-extension or $S$-extension.
If in addition the constants remain unchanged, i.e., $\const{\AR}{\sigma}=\const{\BRSEE}{\sigma}$ an $A$-monomial is also called \rE-monomial, a $P$-monomial is called a \piE-monomial and an $S$-monomial is called a \sigmaE-monomial. In particular, such an $APS$-extension (or $A$-extension or $P$-extension or $S$-extension) is called an \rpisiE-extension (or \rE-extension or \piE-extension or \sigmaE-extension). \\
For the \rpisiE-extension $\dfield{\BRSEE}{\sigma}$ of
$\dfield{\AR}{\sigma}$ we also will write
$\BRSEE=\AR\lr{t_1}\dots\lr{t_e}$. Depending on the case whether
$t_i$ with $1\leq i\leq e$ is an \rE-monomial, \piE-monomial or
\sigmaE-monomial, $\GG\lr{t_i}$ with
$\GG=\AR\lr{t_1}\dots\lr{t_{i-1}}$ stands for the algebraic ring
extension $\GG[t_i]$ with $t_i^n$ for some $n>1$, for the ring of
Laurent polynomials $\GG[t_1,t_1^{-1}]$ or for the polynomial ring
$\GG[t_i]$, respectively.
\end{mdefinition}

\noindent We will rely heavily on the following property of \rpisiE-extensions~\cite[Thm~2.12]{DR1} generalizing the \pisiE-field results given in~\cite{Karr:85,Schneider:01}.

\begin{mtheorem}\label{Thm:RPSCharacterization}
Let $\dfield{\AR}{\sigma}$ be a difference ring. Then
the following holds.
\begin{enumerate}
\item Let $\dfield{\AR[t]}{\sigma}$ be an $S$-extension
of $\dfield{\AR}{\sigma}$ with $\sigma(t)=t+\beta$ where $\beta\in\AR$ such that $\const{\AR}{\sigma}$ is a field. Then this is a \sigmaE-extension (i.e.,
$\const{\AR[t]}{\sigma}=\const{\AR}{\sigma}$) iff there does not exist a
$g\in\AR$ with $\sigma(g)=g+\beta$.

\item Let $\dfield{\AR\ltr{t}}{\sigma}$ be a $P$-extension of
$\dfield{\AR}{\sigma}$ with $\sigma(t)=\alpha\,t$ where $\alpha\in\AR^*$. Then
this is a
\piE-extension (i.e., $\const{\AR\ltr{t}}{\sigma}=\const{\AR}{\sigma}$) iff
there are no $g\in\AR\setminus\{0\}$ and $m\in\ZZ\setminus\{0\}$
with
$\sigma(g)=\alpha^m\,g$.

\item Let $\dfield{\AR[t]}{\sigma}$ be an $A$-extension of $\dfield{\AR}{\sigma}$ of order $\lambda>1$ with $\sigma(t)=\alpha\,t$
where $\alpha\in\AR^*$. Then this is an \rE-extension (i.e.,
$\const{\AR[t]}{\sigma}=\const{\AR}{\sigma}$) iff there are no
$g\in\AR\setminus\{0\}$ and $m\in\{1,\dots,\lambda-1\}$ with
$\sigma(g)=\alpha^m\,g$.
\end{enumerate}
\end{mtheorem}

\noindent In the following we will focus on the special class of simple \rpisiE-extensions.

\begin{mdefinition}
Let $\dfield{\AR}{\sigma}$ be a difference ring extension of $\dfield{\GG}{\sigma}$. Then an \rpisiE-extension $\dfield{\AR\lr{t_1}\dots\lr{t_e}}{\sigma}$ of $\dfield{\AR}{\sigma}$ is called $\GG$-simple if for any \rE-monomial $t_i$ with $1\leq i\leq e$ we have that $\frac{\sigma(t_i)}{t_i}\in(\const{\GG}{\sigma})^*$, and for any \piE-monomial $t_i$ with $1\leq i\leq e$ we have that $\frac{\sigma(t_i)}{t_i}\in\GG^*$. If $\AR=\GG$, we just say simple and not $\GG$-simple.
\end{mdefinition}

\noindent Take a simple \rpisiE-extension
$\dfield{\AR\lr{\mathfrak{t}_1}\dots\lr{\mathfrak{t}_e}}{\sigma}$ of
$\dfield{\AR}{\sigma}$. If $\mathfrak{t}_i$ is an \rE-monomial or
\piE-monomial, we can reorder the generator and obtain the
difference ring $\dfield{\BRSEE}{\sigma}$ with
$\BRSEE=\AR\lr{\mathfrak{t}_i}\lr{\mathfrak{t}_1}\dots\lr{\mathfrak{t}_{i-1}}\lr{\mathfrak{t}_{i+1}}\dots\lr{\mathfrak{t}_e}$.
Note that this rearrangement does not change the set of constants.
Further note that the recursive nature of $\sigma$ is respected
accordingly. Thus $\dfield{\BRSEE}{\sigma}$ is again a simple
\rpisiE-extension of $\dfield{\AR}{\sigma}$. Applying such
permutations several times enables one to move all \rE-monomials and
\piE-monomials to the left and all the \sigmaE-monomials to the
right yielding a simple \rpisiE-extension of the form
$\dfield{\AR\lr{t_1}\dots\lr{t_u}\lr{\tau_1}\dots\lr{\tau_v}}{\sigma}$
with $u+v=e$ where the $t_i$ with $1\leq i\leq u$ are \rE- or
\piE-monomials and the $\tau_i$ with $1\leq i\leq v$ are
\sigmaE-monomials.

\medskip

We emphasize that the class of simple \rpisiE-extensions defined over the rational difference field, $q$-rational difference or mixed multibasic difference field (see Example~\ref{Exp:BaseFields}) cover
all the indefinite nested summation objects that the we have encountered so far in practical problem solving: this class enables one to treat $(-1)^k$, e.g., with the \rE-monomial $t_1$ with $\sigma(t_1)=-t_1$ and $t_1^2=1$ and more generally it allows one to formulate all hypergeometric/$q$-hypergeometric/mixed multibasic hypergeometric products~\cite{Schneider:05c,OS:17} and nested sums defined over such products~\cite{DR1,DR2}.
For the definition of the hypergeometric class see Definition~\ref{Def:IndefiniteNested} below.
In particular, this class enables one to represent d'Alembertian solutions~\cite{Abramov:94,Abramov:96}, a subclass of Liouvillian solutions~\cite{Singer:99,Petkov:2013}.

For \pisiE-fields $\dfield{\GG}{\sigma}$ and more generally for
simple \rpisiE-extensions $\dfield{\AR}{\sigma}$ of
$\dfield{\GG}{\sigma}$ many variations of Problem~RPT have been
worked out~\cite{Karr:81,DR1,DR2,DR3}. In this regard, the depth
function $\depth:\AR\to\NN$ will be used. More precisely, let
$\dfield{\AR}{\sigma}$ be a simple \rpisiE-extension of
$\dfield{\GG}{\sigma}$ with $\AR=\GG\lr{t_{1}}\dots\lr{t_e}$. By
definition we have $\sigma(t_i)=\alpha_i\,t_i+\beta_i$ for $1\leq
i\leq e$ where the $\alpha_i,\beta_i$ are taken from the ring below.
Then the depth function $f:\AR\to\NN$ is defined iteratively as
follows. For $f\in\GG$ we set $\depth(f)=0$. If $\depth$ has been
defined for $\BRSEE=\AR\lr{t_1}\dots\lr{t_{i-1}}$, then
define\footnote{For a finite set $L\subseteq\AR$ we define
$\text{supp}(L)=\{1\leq j\leq e\mid t_j\text{ occurs in }L\}$.}
$\depth(t_i)=1+\max_{j\in\text{supp}(\{\alpha_i,\beta_i\})}\depth(\tau_j)$.
Further, for $f\in\BRSEE\lr{t_i}$, define
$\depth(f)=\max_{j\in\text{supp}(f)}\depth(t_j)$. In other words
$\depth(t_i)$ for $1\leq i\leq e$ gives the maximal nesting depth of
an \rpisiE-monomial and $\depth(f)$ for $f\in\AR$ measures the
maximal nesting depth among all the arising \rpisiE-monomials $t_i$
in $f$.

In the following we emphasize the following four variants of Problem~RPT that will play a role in this article.

\begin{mremark}\label{Remark:RPTVariants}
Let $\dfield{\AR}{\sigma}$ be a simple \rpisiE-extension of a \pisiE-field\footnote{In order to apply our summation algorithms, we must assume that the constant field $\KK=\const{\GG}{\sigma}$ has certain algorithmic properties~\cite{Schneider:05c}; this is guaranteed if we are given, e.g., a rational function field $\KK=\KK'(x_1,\dots,x_l)$ over an algebraic number field $\KK'$.} $\dfield{\GG}{\sigma}$ with $\AR=\GG\lr{t_{1}}\dots\lr{t_e}$ and let $f_1,\dots,f_d\in\AR$. Then the following strategies are proposed that can be executed within the summation package \texttt{Sigma}.

\begin{description}
 \item[$\text{RPT}_1$:] Decide constructively, if Problem RPT is solvable with $\BRSEE=\AR$; see~\cite{DR1}.


 \item[$\text{RPT}_2$:]  Try to solve Problem $\text{RPT}_1$. If this is not possible, decide constructively if there is a \sigmaE-extension $\dfield{\BRSEE}{\sigma}$ of $\dfield{\AR}{\sigma}$ with $\BRSEE=\AR[\tau_1]\dots[\tau_v]$ in which one finds the desired $c_i$ and $g\in\BRSEE$ with the extra property that $\depth(\tau_i)\leq\depth(c_1\,f_1+\dots+c_d\,f_d)$ holds for all $1\leq i\leq v$; see~\cite{Schneider:08} in combination with~\cite{DR1}.

\item[$\text{RPT}_3$:]  By the recursive nature, we may reorder the \rpisiE-monomials in $\AR$ such that
$\depth(t_1)\leq\depth(t_2)\leq\dots\leq\depth(t_e)$ holds. With this preparation step, try to solve Problem~$\text{RPT}_2$. If this is not possible,
decide if there is a \sigmaE-extension
$\dfield{\BRSEE}{\sigma}$ of $\dfield{\AR}{\sigma}$ with
$\BRSEE=\AR[\tau]$ and
$\sigma(\tau)-\tau\in\GG\lr{t_1}\dots\lr{t_i}$ for some $0\leq i<e$
where at least one of the $t_{i+1},\dots,t_e$ occurs in
$c_1\,f_{1}+\dots+c_d\,f_{d}$ and $i$ is minimal among all such
possible choices; see~\cite{Schneider:14} in combination
with~\cite{DR1}.


\item[$\text{RPT}_4$:] Try to solve Problem $\text{RPT}_3$. If this is not possible, it follows in particular that there is no $g\in\AR$ with $\sigma(g)-g=f_1$. By part~1 of Theorem~\ref{Thm:RPSCharacterization} we can construct the \sigmaE-extension $\dfield{\AR[\tau]}{\sigma}$ of $\dfield{\AR}{\sigma}$ with $\sigma(\tau)=\tau+f_1$ and return $g=\tau$ and $c_1=1$, $c_i=0$ for $2\leq i\leq d$.


\end{description}
\end{mremark}

\begin{mexample}\label{Exp:PiSi}
Consider the definite sum
$X(k)=\sum_{j=0}^k F(k,j)$ with the summand $F(k,j)=\binom{k}{j} S_1({j})^2$ and the shifted versions
$F(k+i,j)=\prod_{l=1}^i\frac{k + l}{k - j + l}\binom{k}{j} S_1({j})^2$ for $i=0,1,2,\dots$. We start with the difference field $\dfield{\KK(t)}{\sigma}$ with $\sigma(t)=t+1$ and constant field $\KK=\QQ(k)$. Further we rephrase $\tbinom{k}{j}$ with the shift behavior $\tbinom{k}{j+1}=\frac{k-j}{j+1}\,\tbinom{k}{j}$ with $b$ in the \piE-extension $\dfield{\KK(t)[b,b^{-1}]}{\sigma}$ of $\dfield{\KK(t)}{\sigma}$ with $\sigma(b)=\frac{k-t}{t+1}b$. Further, we rephrase $S_1(k)$ with $h$ in the \sigmaE-extension $\dfield{\KK(t)[b,b^{-1}][h]}{\sigma}$ of $\dfield{\KK(t)[b,b^{-1}]}{\sigma}$ with $\sigma(h)=h+\frac1{t+1}$.
In this ring we are now in the position to represent $F_i(j)=F(k+i-1,j)$ with $f_{i}=\prod_{l=1}^{i-1}\frac{k + l}{k - j + l}b\,h^2$ for $i=1,2,\dots$. First we will solve Problem~RPT with the simplest variant $\text{RPT}_1$. We start with $d=0,1,...$ and are successful with $d=5$: \texttt{Sigma} computes
$c_1=-8 (1+k) (3+k)$, $c_2= 4 \big(
                29+25 k+5 k^2\big)$, $c_3=-2 (8+3 k) (10+3 k)$, $c_4= 86+49 k+7 k^2$, $c_5=-(4+k)^2$
and
\begin{align*}
g=&b \big(
                \tfrac{(1+k) (2+k) (3+k)}{(1
                -t
                +k
                ) (2
                -t
                +k
                ) (3
                -t
                +k
                )}
                -\tfrac{2(1+k)(
                        10 t
                        -6 t^2
                        +6 t k
                        -2 t^2 k
                        +t k^2
                )}{(1
                -t
                +k
                ) (2
                -t
                +k
                ) (3
                -t
                +k
                )}h\\
                &+\tfrac{ t^2 (1+k) \big(
                        56
                        -56 t
                        +12 t^2
                        +42 k
                        -30 t k
                        +4 t^2 k
                        +11 k^2
                        -4 t k^2
                        +k^3
                \big)}{(-4
                +t
                -k
                ) (-3
                +t
                -k
                ) (-2
                +t
                -k
                ) (-1
                +t
                -k
                )}h^2\big)
\end{align*}
such that~\eqref{Equ:ParaDF} holds. Reinterpreting $b$ and $h$ as $\binom{k}{j}$ and $S_1(j)$, yields a solution of~\eqref{eq:ptcert} for $F_i(j)$. Finally, summing this relation over $j$ from $0$ to $k$ and taking care of compensating terms yields the linear recurrence relation
\begin{multline}\label{Equ:XOrder4}
X(4+k)=-\tfrac{8 (1+k) (3+k)}{(4+k)^2}X(k)
+\tfrac{4(
        29+25 k+5 k^2)}{(4+k)^2}X(1+k)\\
-\tfrac{2 (8+3 k) (10+3 k)}{(4+k)^2}X(2+k)
+\tfrac{86+49 k+7 k^2}{(4+k)^2}X(3+k)+\tfrac{1}{(4+k)^2}.
\end{multline}
$\text{RPT}_2$ will not contribute to a shorter recurrence. However, applying $\text{RPT}_3$ \texttt{Sigma} finds for $d=2$ the solution
$c_1=-4 (1+k)$, $c_2=2 (3+2 k)$, $c_3=-2-k$ and
$$g=-(1+k) (2+k)\tau
+\big(
        -\tfrac{2 (1+k)}{-1
        +t
        -k
        } h
        +\tfrac{t (1+k) (-2
        +2 t
        -k
        )}{(-2
        +t
        -k
        ) (-1
        +t
        -k
        )} h^2
\big) b$$
within the \sigmaE-extension $\dfield{\KK(t)[b,b^{-1}][h][\tau]}{\sigma}$ of $\dfield{\KK(t)[b,b^{-1}][h]}{\sigma}$ with $\sigma(\tau)=\tau+\frac{b}{(1+t)^2 (1
+k
-t
)}$. Interpreting $\tau$ as the sum $s(k,j)=\sum_{i=1}^j \frac{\binom{k}{i}}{i (1
-i
+k
) (2
-i
+k
)}$ and performing similar steps as above, \texttt{Sigma} produces the linear recurrence
\begin{multline}\label{Equ:XOrder2Raw}
-4 (1+k) X(k)
+2 (3+2 k) X(1+k)
+(-2-k) X(2+k)\\
=\tfrac{-5-5 k-k^2}{(1+k) (2+k)}
-(1+k) (2+k)
\sum_{i=1}^k \tfrac{\binom{k}{i}}{i (1
-i
+k
) (2
-i
+k
)}.
\end{multline}
Note that the sum $s(k)=s(k,k)$ on the right hand side is given in a form that is not indefinite nested and thus cannot be represented automatically in terms of an \rpisiE-extension. However, the sum $s(k)$ can be simplified further. Applying  $\text{RPT}_1$ \texttt{Sigma} computes the linear recurrence
$$-2 (1+k) (2+k) s(k)
+(2+k) (10+3 k) s(1+k)
-(4+k)^2 s(2+k)
=\tfrac{4-k-k^2}{(1+k) (2+k) (3+k)}$$
and solves the recurrence in terms of d'Alembertian solutions:
$$\{\tfrac{c_1}{(1+k) (2+k)} +c_2\Big[\tfrac{(4+3 k) 2^{-2+k}}{(1+k)^2 (2+k)^2}
        +\tfrac{S_1({{2},k})}{8 (1+k) (2+k)}\Big] -\tfrac{2 (3+2 k)}{(1+k)^2 (2+k)^2}
        -\tfrac{S_1({k})}{(1+k) (2+k)}\mid c_1,c_2\in\KK\}.$$
Finally, taking the two initial values $s(1)=\frac12$ and $s(2)=\frac7{12}$ determines $c_1=-1$ and $c_2=8$ so one gets
\begin{equation}\label{Equ:RHSSum}
s(k)=-\tfrac{8+7 k+k^2}{(1+k)^2 (2+k)^2}
+\tfrac{(4+3 k) 2^{1+k}}{(1+k)^2 (2+k)^2}
-\tfrac{1}{(1+k) (2+k)}S_1({k})
+\tfrac{1}{(1+k) (2+k)}S_1({{2},k}).
\end{equation}
Thus recurrence~\eqref{Equ:XOrder2Raw} can be simplified to
\small
\begin{multline}\label{Equ:XOrder2}
X(2+k)=
-\tfrac{4 (1+k)}{2+k}X(k)
+\tfrac{2 (3+2 k)}{2+k}X(1+k)
+\tfrac{-3-2 k}{(1+k) (2+k)^2}
+\tfrac{2^{1+k} (4+3 k)}{(1+k) (2+k)^2}
+\tfrac{S_1({k})}{-2-k}
-\tfrac{S_1({{2},k})}{-2-k};
\end{multline}
\normalsize
we emphasize that the sums in the inhomogeneous part of~\eqref{Equ:XOrder2} are now all indefinite nested and can be rephrased in an \rpisiE-extension.
We note that we can solve this recurrence (or alternatively the recurrence~\eqref{Equ:XOrder4}) again by solving the recurrence in terms of d'Alembertian solutions using \texttt{Sigma}. This finally enables us to find the closed form representation
\begin{multline}\label{Equ:XOrder0}
X(k)= 2^k\big(
        -2 S_1({k}) S_1\big({{\tfrac{1}{2}},k}\big)
        -2 S_2\big({{\tfrac{1}{2}},k}\big)
        +3 S_{1,1}\big({{\tfrac{1}{2},1},k}\big)\\
        -S_{1,1}\big({{\tfrac{1}{2},2},k}\big)
        +S_1({k})^2
        +S_2({k})
\big).
\end{multline}
\end{mexample}

Later we will consider the sum  $S(n)=\sum_{k=0}^n \binom{n}{k} X(k)$ with $X(k)=\sum_{j=0}^k \binom{k}{j} S_1({j})^2$ and aim at computing a linear recurrence in $n$. One option is to take the representation~\eqref{Equ:XOrder0} and to use one of the tactics from Remark~\ref{Remark:RPTVariants} -- this is our usual strategy from~\cite{Schneider:13} to tackle such sums.
In Example~\ref{Exp:SimpleX} below we will follow an alternative strategy. Instead of working with the zero-order recurrence~\eqref{Equ:XOrder0} we will work with the higher-order recurrences~\eqref{Equ:XOrder4} or~\eqref{Equ:XOrder2}. The advantage will be to work in a smaller \rpisiE-extension and encoding parts of the expression~\eqref{Equ:XOrder0} within the recurrence operator. In order to accomplish this new strategy, we introduce and explore higher-order extensions in the next subsection.

\subsection{Higher order linear extensions}

So far we have considered indefinite nested sums of the form $S(k)=\sum_{i=l}^kF(i)$ and products of the form $P(k)=\prod_{i=l}^kF(i)$ with $l\in\NN$ that can be encoded by the first order homogeneous recurrences
$S(k+1)=S(k)+F(k+1)$ and $P(k+1)=F(k+1)\,P(k)$, respectively.
However, many interesting summation objects can be only described by higher-order recurrences, like Legendre polynomials, Hermite polynomials, or Bessel functions. More precisely, we are interested in dealing with a sequence $X(k)$ which satisfies a linear recurrence
\begin{equation}\label{Equ:HolonomicRec}
X(k+s+1)=A_0(k)\,X(k)+A_1(k)\,X(k+1)+\dots+A_s(k)\,X(k+s)+A_{s+1}(k)
\end{equation}
where the sequences $A_i(k)$ ($1\leq i\leq s+1$) are expressible in a difference ring $\dfield{\AR}{\sigma}$.

\begin{mremark}
Sequences that satisfy~\eqref{Equ:HolonomicRec} are also called holonomic. Specializing to the case that the $A_i(k)$ with $0\leq i\leq s+1$ are elements of the rational or $q$-rational difference field, many important properties have been elaborated~\cite{Zeilberger:90a,Salvy:94,Mallinger,Chyzak:98b,CHYZAK2000115,KK:09,KP:11}.
\end{mremark}

\noindent Then the summation object $X(k)$ with the recurrence relation~\eqref{Equ:HolonomicRec} can be represented in a higher order difference ring extension as follows~\cite{Schneider:05a}.

\begin{mdefinition}
A \emph{higher-order linear difference ring extension} (in short
\emph{h.o.l.\ extension}) $\dfield{\BRSEE}{\sigma}$ of a difference
ring $\dfield{\AR}{\sigma}$ is a polynomial ring extension
$\BRSEE=\AR[x_0,\dots,x_s]$ with the variables $x_0,\dots,x_s$ and
the automorphism $\sigma:\BRSEE\to\BRSEE$ is extended from $\AR$ to
$\BRSEE$ subject to the relations $\sigma(x_i)=x_{i+1}$ for $0\leq
i<s$ and
\begin{equation}\label{HOEShift}
\sigma(x_s)=a_0\,x_0+a_1\,x_1+\dots+a_{s}\,x_{s}+a_{s+1}
\end{equation}
for some $a_0,\dots,a_{s+1}\in\AR$. $s+1$ is also called the extension order or recurrence order.
\end{mdefinition}

\noindent Namely, if we rephrase $X(k)$ as $x_0$, then $X(k+1)$ corresponds to $\sigma(x_0)=x_1$, $X(k+2)$ corresponds to $\sigma(x_1)=x_2$, etc. Finally, $X(k+s)$ corresponds to $x_s$ and the relation~\eqref{Equ:HolonomicRec} is encoded by~\eqref{HOEShift}.

Concerning concrete computations, we usually start with a \pisiE-field $\dfield{\GG}{\sigma}$ (in particular as defined in Example~\ref{Exp:BaseFields}) in which the $A_i(k)$ are encoded by $a_i\in\GG$. Further, we assume that $A_{s+1}(k)$ can be rephrased as $a_{s+1}$ in an \rpisiE-extension $\dfield{\AR}{\sigma}$ of $\dfield{\GG}{\sigma}$.
Then we construct the h.o.l.\ extension $\dfield{\HH}{\sigma}$ of $\dfield{\AR}{\sigma}$ with $\HH=\AR[x_0,\dots,x_s]$ and~\eqref{HOEShift} with $a_0,\dots,a_{s}\in\GG$ and $a_{s+1}\in\AR$.\\
In the following we will work out summation algorithms that tackle
Problem~RPS in $\dfield{\HH}{\sigma}$ with $f_i\in
\AR\,x_0+\dots+\AR\,x_s+\AR$ for $1\leq i\leq d$ and
$g\in\BRSEE\,x_0+\dots+\BRSEE\,x_s+\BRSEE$ for an appropriate
difference ring extension $\dfield{\BRSEE}{\sigma}$ of
$\dfield{\AR}{\sigma}$. To warm up we will first focus on the
following basic telescoping problem.

\medskip

\noindent\textit{Given} $f\in\HH$ with $f=f_0\,x_0+\dots+f_s\,x_s+f_{s+1}$ where $f_i\in\AR$ for $0\leq i\leq s+1$.\\
\textit{Find}, if possible, $g=g_0\,x_0+\dots+g_s\,x_s+g_{s+1}$ where $g_i\in\AR$ for $0\leq i\leq s+1$.

\medskip

\noindent In order to tackle this problem (and more generally Problem~RPT below) we rely on the following lemma that follows immediately by~\cite[Lemmas~1,2]{Schneider:05a}; the proof is based on coefficient comparison.

\begin{mlemma}\label{Lemma:IndefiniteConstraint}
Let $\dfield{\AR[x_0,\dots,x_s]}{\sigma}$ be a h.o.l.\ extension of $\dfield{\AR}{\sigma}$ with~\eqref{HOEShift}. Let
$f=f_{0}\,x_0+\dots+f_{s}\,x_s+f_{s+1}$ with $f_{i}\in\AR$ and $g=g_0\,x_0+\dots+x_s\,x_s+g_{s+1}$ with $g_i\in\AR$. Then
\begin{equation}\label{Equ:RefinedTele}
\sigma(g)-g=f
\end{equation}
if and only if the following equations hold:
\begin{align}\label{Equ:HigherRec}
\sum_{j=0}^s\sigma^{s-j}(a_j)\sigma^{s-j+1}(g_s)-g_s=\sum_{j=0}^s\sigma^{s-j}(f_{j}),\\
\sigma(g_{s+1})-g_{s+1}=f_{s+1}-a_{s+1}\,\sigma(g_s),\label{Equ:InHomTele}
\end{align}
\begin{align}
g_0&=a_0\,\sigma(g_s)-f_0,\label{Equ:Computeg0}\\
g_i&=\sigma(g_{i-1})+a_i\,\sigma(g_s)-f_i,\quad(0\leq i<s).\label{Equ:Computegi}
\end{align}
\end{mlemma}

\noindent Namely, suppose that we succeed in computing $g_s\in\AR$ and $g_{s+1}\in\AR$ with~\eqref{Equ:HigherRec} and~\eqref{Equ:InHomTele}. Then we can compute $g_0,\dots,g_{s-1}\in\AR$ using~\eqref{Equ:Computeg0} and~\eqref{Equ:Computegi}, and by Lemma~\ref{Lemma:IndefiniteConstraint} it follows that $g=g_0\,x_0+\dots+x_s\,x_s+g_{s+1}$ is a solution of~\eqref{Equ:RefinedTele}.

\begin{mremark}\label{Remark:ZeroOrder}
If $s=0$, constraint~\eqref{Equ:HigherRec} is nothing else than $\sigma(g_s)-g_s=0$ which gives the solution $g_s=1$. Hence what remains is constraint~\eqref{Equ:InHomTele} which reduces to $\sigma(g_{s+1})-g_{s+1}=f_{s+1}$. In other words, in this special case Lemma~\ref{Lemma:IndefiniteConstraint} boils down to the telescoping problem in $\dfield{\AR}{\sigma}$.
\end{mremark}

\begin{mexample}
Consider the sum
$S(n)=\sum_{k=0}^n F(k)$ with the summand $F(k)=\frac{X(k)}{2^k}$
where the sequence $X(k)$ is determined by the recurrence
\begin{equation}\label{Equ:XOrder2Triv}
X(2+k)=
-\tfrac{4 (1+k)}{2+k}X(k)
+\tfrac{2 (3+2 k)}{2+k}X(1+k)-\tfrac{1}{2+k}
\end{equation}
and the initial values $X(0)=0$, $X(1)=-1$. We take the rational difference field $\dfield{\GG}{\sigma}$ with $\GG=\QQ(t)$ and $\sigma(t)=t+1$ and construct the \piE-extension $\dfield{\AR}{\sigma}$ of $\dfield{\GG}{\sigma}$ with $\AR=\GG\ltr{p}$ and $\sigma(p)=2\,p$. Finally, we construct the h.o.l.\ extension $\dfield{\HH}{\sigma}$ of $\dfield{\AR}{\sigma}$ with
$\sigma(x_2)=-\frac{4 (1+t)}{2+t}x_0
+\frac{2 (3+2 t)}{2+t}x_1-\frac1{2+t}$
and search for $g=g_0\,x_0+g_1\,x_1+g_2$ with $g_i\in\AR$ such that $\sigma(g)-g=\frac{x_0}{p}$ holds.
The constraint~\eqref{Equ:HigherRec} of Lemma~\ref{Lemma:IndefiniteConstraint} reads as
$-\frac{4 (2+t)}{3+t}\sigma^2(g_1)+\frac{2 (3+2 t)}{2+t}\sigma(g_1)-g_1=\frac{1}{2 p}.$
Using \texttt{Sigma} we compute
$g_1=-\frac{(-1+t) (1+t)}{2 p}\in\AR$ and we get the constraint (compare~\eqref{Equ:InHomTele}):
$\sigma(g_2)-g_2=-\frac{t}{4 p}.$
Solving this telescoping equation gives $g_2=\frac{1+t}{2\,p}\in\AR$.
Further, using~\eqref{Equ:Computeg0}, we obtain
$g_0=\frac{-1+t+t^2}{p}$. Reinterpreting $g$ in terms of our summation objects yields
$G(k)=
        \frac{-1+k+k^2}{2^{k}} X(k)
        -\frac{(-1+k) (1+k)}{2\,2^{k}} X(k+1)+\frac{1+k}{2\,2^{k}}$
with
$\frac{X(k)}{2^k}=G(k+1)-G(k)$.
Finally, summing this equation over $k$ from $0$ to $n$ produces
$$S(n)=\frac{1+n}{2^{1+n}}\Big(1
+2 \,n\, X(n)+(1-n)\,X(1+n)\Big).$$
\end{mexample}

\begin{mremark}
More generally, multivariate sequences are often described by a system of homogeneous linear recurrences with coefficients from the difference field $\dfield{\KK(t)}{\sigma}$ with $\sigma(t)=t+1$ or with $\sigma(t)=q\,t$. Then the telescoping problem with $\AR=\KK(t)$ and more generally, the parameterized telescoping problem,  can be treated in this setting using the holonomic system approach~\cite{Zeilberger:90a}. In this regard, Chyzak's fast algorithm~\cite{CHYZAK2000115} was a major breakthrough that has been improved further in~\cite{Koutschan:10}. Lemma~\ref{Lemma:IndefiniteConstraint} specializes to one linear recurrence (and does not treat a system in the multivariate sequence case). However, it dispenses the user to work with Gr\"obner bases and expensive uncoupling procedures~\cite{Zuercher:94,BCP13} that are needed in the standard approaches~\cite{CHYZAK2000115,Koutschan:10}. In particular, the constraints~\eqref{Equ:HigherRec} and~\eqref{Equ:Computeg0} have been worked out explicitly which will be the basis for further explorations. An extra bonus is the treatment of inhomogeneous recurrences that will be utilized below.\\
We remark further that a special case of Lemma~\ref{Lemma:IndefiniteConstraint} can be also related to~\cite{Abramov:99}.
\end{mremark}

Suppose that the summand $F(k)$ can be rephrased by $f$ in a difference ring $\dfield{\AR}{\sigma}$ as constructed above.
In most applications, one will fail to find a telescoping solution for $f$ in $\AR$. To gain more flexibility, we will consider two strategies.
\begin{enumerate}
\item[(I)] Try to extend the difference ring $\dfield{\AR}{\sigma}$ with a simple \rpisiE-extension in which one finds a telescoping solution.
\item[(II)] In case that the summand $F(k)$ contains an extra parameter, say $F(k)=F(n,k)$, utilize the creative telescoping paradigm with $F_i(k)=F(n+i-1,k)$ for $1\leq i\leq d$.
\end{enumerate}

\noindent As it turns out below, the successful application of strategy I can be connected to the problem of finding constants in a difference ring or equivalently to construct higher-order extensions with smaller recurrence order. In Subsection~\ref{Sec:FindingConstants} we will provide a constructive theory that enables one compute such constants and thus to find improved higher order extensions in the setting of simple \rpisiE-extensions. Based on this insight, we will propose in Subsection~\ref{Subsec:RPTHolo} our algorithm to solve the parameterized telescoping problem in $\dfield{\AR}{\sigma}$ or in a properly chosen simple \rpisiE-extension of it. In a nutshell, we will combine strategies (I) and (II) that will lead to efficient and flexible algorithms to tackle indefinite and definite summation problems.

\subsubsection{Finding constants or finding recurrences with lower order}\label{Sec:FindingConstants}

We are interested in the following problem.

\begin{center}
 \noindent\fbox{\begin{minipage}{11.4cm}
\noindent \textbf{Problem C for $\dfield{\AR}{\sigma}$:} Find a linear constant.\\
\small
\textit{Given} a h.o.l.\ extension  $\dfield{\AR[x_0,\dots,x_s]}{\sigma}$ of $\dfield{\AR}{\sigma}$ with~\eqref{HOEShift} where $a_0,\dots,a_{s+1}\in\AR$ .\\
\textit{Find} a $g=g_0\,x_0+\dots+g_s\,x_s+g_{s+1}\in\HH\setminus\AR$ with $g_i\in\AR$ (or in an appropriate extension of it) such that $\sigma(g)=g$ holds.
\end{minipage}}
\end{center}

\noindent Setting $f_i=0$ for $0\leq i\leq s+1$ in Lemma~\ref{Lemma:IndefiniteConstraint} yields a basic strategy for Problem~C.

\begin{mlemma}\label{Lemma:FindConstantsInside}
Let $\dfield{\HH}{\sigma}$ be a h.o.l.\ extension of $\dfield{\AR}{\sigma}$ with $\HH=\AR[x_0,\dots,x_s]$ and~\eqref{HOEShift} where $a_i\in\AR$. Then there exists a $g=g_0\,x_0+\dots+g_s\,x_s+g_{s+1}\in\HH\setminus\AR$ with $g_i\in\AR$ such that $\sigma(g)=g$
if and only if there is a $g\in\AR\setminus\{0\}$ with
\begin{equation}\label{Equ:DEHomConst}
\sum_{j=0}^s\sigma^{s-j}(a_j)\sigma^{s-j+1}(h)-h=0
\end{equation}
and a $\gamma\in\AR$ with
\begin{equation}\label{Equ:DEInhomConst}
\sigma(\gamma)-\gamma=-a_{s+1}\,\sigma(h).
\end{equation}
In this case we can set $g_s=h$, $g_{s+1}=\gamma$, $g_0=a_0\,\sigma(g_s)$ and $g_i=\sigma(g_{i-1})+a_i\,\sigma(h)$ for $1\leq i< s$.
\end{mlemma}

\noindent In other words, if one finds an $h\in\AR\setminus\{0\}$ with \eqref{Equ:DEHomConst} and a $\gamma\in\AR$ with~\eqref{Equ:DEInhomConst}, one can compute a $g\in\HH\setminus\AR$ with $\sigma(g)=g$.

\medskip

\noindent \textbf{From constants to recurrences with smaller order.} Now suppose that we find such a $g\in\HH\setminus\AR$ with $\sigma(g)=g$ and reinterpret $g$ as
$$G(k)=G_0(k)X(k)+\dots+G_s(k)X(k+s)+G_{s+1}(k)$$
where we rephrase for $1\leq i\leq s+1$ the $g_i$ in terms of our summation objects yielding the expression $G_i(k)$.
Suppose that $G(k+1)=G(k)$ holds for all $k\in\NN$ with $k\geq\lambda$ for some $\lambda$ chosen big enough.
Evaluating $c:=G(\lambda)\in\KK$ with our given sequence $X(k)$ gives the identity $G(k)=c$. In other words, we find the new linear recurrence
\begin{equation}\label{Equ:ShorterXRec}
G_0(k)X(k)+\dots+G_s(k)X(k+s)=c-G_{s+1}(k)
\end{equation}
with order $s$; note that so far we used the recurrence~\eqref{Equ:HolonomicRec} to model the object $X(k)$ which has order $s+1$.
Now suppose that $g_s\in\AR^*$ holds. Then we can define the h.o.l.\ extension $\dfield{\AR[y_0,\dots,y_{s-1}]}{\sigma}$ of $\dfield{\AR}{\sigma}$ with $\sigma(y_{s-1})=g'_0\,y_0+\dots+g'_{s-1}\,y_{s-1}+g'_s$ with $g'_i=-\frac{g_i}{g_{s}}$ for $0\leq i<s$ and $g'_s=\frac{c-g_{s+1}}{g_s}$.

\medskip

\noindent Summarizing, finding a constant indicates that the recurrence used to describe the object $X(k)$ is not optimal. But given such a constant also enables one to cure the situation. One can derive a recurrence that models $X(k)$ with a smaller order. Before we look at a concrete application in Example~\ref{Exp:ConstantReduction} below, we will work out the different possible scenarios to hunt for constants. So far we considered

\medskip

\noindent\textbf{Case 1.1:} there is an $h\in\AR\setminus\{0\}$ with~\eqref{Equ:DEHomConst} and a $\gamma\in\AR$ with~\eqref{Equ:DEInhomConst}. Then we activate Lemma~\ref{Lemma:FindConstantsInside} and find
\begin{equation}\label{Equ:ConstgInA}
g=g_0\,x_0+\dots+g_s\,x_s+g_{s+1}
\end{equation}
with $g_i\in\AR$ for all $0\leq i\leq s+1$, $g_s\neq0$, with $\sigma(g)=g$.

\medskip

It might happen that one finds an $h\in\AR\setminus\{0\}$ with~\eqref{Equ:DEHomConst} but one fails to find a $\gamma\in\AR$ with~\eqref{Equ:DEInhomConst}. This situation can be covered as follows.

\medskip

\noindent\textbf{Case~1.2:} There is no $\gamma\in\AR$ with~\eqref{Equ:DEInhomConst}. By  part~(1) of Theorem~\ref{Thm:RPSCharacterization} we can construct a \sigmaE-extension $\dfield{\AR[\tau]}{\sigma}$ of $\dfield{\AR}{\sigma}$ with $\sigma(\tau)=\tau-a_{s+1}\,\sigma(h)$ and can put on top our h.o.l.\ extension $\dfield{\AR[\tau][x_0,\dots,x_s]}{\sigma}$ of $\dfield{\AR[\tau]}{\sigma}$ with~\eqref{HOEShift}. Then by Lemma~\ref{Lemma:FindConstantsInside} we get
\begin{equation}\label{Equ:ConstgInSigma}
g=g_0\,x_0+\dots+g_s\,x_s+\tau
\end{equation}
with $g_i\in\AR$ for $0\leq i\leq s$ such that $\sigma(g)=g$ holds.

\medskip

Now let us tackle the case that there is no $h\in\AR$
with~\eqref{Equ:DEHomConst} but there is such an $h$ in a simple
\rpisiE-extension. More precisely, we assume that
$\dfield{\AR}{\sigma}$ itself is a simple \rpisiE-extension of
$\dfield{\GG}{\sigma}$ and that for~\eqref{HOEShift} we have that
$a_0,\dots,a_s\in\GG$ and $a_{s+1}\in\AR$. In this setting, suppose
that there is a $\GG$-simple \rpisiE-extension
$\dfield{\BRSEE}{\sigma}$ of $\dfield{\AR}{\sigma}$ in which one
finds an $h\in\BRSEE$ with~\eqref{Equ:DEHomConst}. Note that
$\dfield{\BRSEE}{\sigma}$ is a simple \rpisiE-extension of
$\dfield{\GG}{\sigma}$. Then we can apply the following result. A
simpler field version can be found
in~\cite[Lemma~4.5.3]{Schneider:01}; for the Liouvillian case with
$\AR=\KK(t)$ and $\sigma(t)=t+1$ we refer
to~\cite[Thm.~5.1]{Singer:99}.

\begin{mproposition}\label{Prop:ConstraintForRecSol}
Let $\dfield{\AR}{\sigma}$ with
$\AR=\GG\lr{\tilde{t}_1}\dots\lr{\tilde{t}_{\tilde{u}}}[\tilde{\tau}_1]\dots[\tilde{\tau}_{\tilde{v}}]$
be a simple \rpisiE-ex\-tension of $\dfield{\GG}{\sigma}$, and let
$\dfield{\BRSEE}{\sigma}$ with
$\BRSEE=\AR\lr{t_1}\dots\lr{t_u}[\tau_1]\dots[\tau_v]$ be a
$\GG$-simple \rpisiE-extension of $\dfield{\AR}{\sigma}$ where the
$\tilde{t}_i,t_i$ are \rpiE-monomials and the
$\tilde{\tau}_i,\tau_i$ are \sigmaE-monomials. Let $f\in\AR$ and
$a_0,\dots,a_s\in\GG$. Suppose there is a $g\in\BRSEE\setminus\{0\}$
with
\begin{equation}\label{Equ:GenInhomSol}
a_s\,\sigma^s(g)+\dots+a_0\,g=f.
\end{equation}
\begin{enumerate}
\item
If $f=0$ or $g\notin\AR\lr{t_1}\dots\lr{t_u}$,
then there are $l_i,\tilde{l}_i\in\ZZ$ and $w\in\GG^*$ such that for $h=w\,{\tilde{t}_1}^{\tilde{l}_1}\dots \tilde{t}_{\tilde{u}}^{\tilde{l}_{\tilde{u}}}t_1^{\tilde{l}_1}\dots t_u^{l_u}$ we have
\begin{equation}\label{Equ:GenHomSol}
a_s\,\sigma^s(h)+\dots+a_0\,h=0.
\end{equation}
\item Otherwise, if $f\neq0$ and $g\in\AR\lr{t_1}\dots\lr{t_u}$, there is also a solution of~\eqref{Equ:GenInhomSol} in $\AR$.
\end{enumerate}
\end{mproposition}
\begin{proof}
(1) Set $\BRSEE_j=\AR\lr{t_1}\dots\lr{t_u}[\tau_1]\dots[\tau_j]$ for
$0\leq j\leq v$. First we show that there is an
$h\in\BRSEE\setminus\{0\}$ with~\eqref{Equ:GenHomSol}. If $f=0$,
this holds by assumption. Otherwise, we can conclude that there is a
$g\in\BRSEE\setminus\BRSEE_0$ with~\eqref{Equ:GenInhomSol} again by
assumption. Now take among all the possible $g$
with~\eqref{Equ:GenInhomSol} an element
$g\in\BRSEE_i\setminus\BRSEE_{i-1}$ where $i>0$ is minimal.
Then $g=h\,\tau_i^m+b$ for some $m>0$ and $h,b\in\BRSEE_{i-1}$ with $h\neq0$. By coefficient comparison w.r.t.\ $\tau_i$ in~\eqref{Equ:GenInhomSol} and using the fact that $\sigma(\tau_i)-\tau_i\in\BRSEE_{i-1}$ and $a_i\in\GG$, we conclude that $h$ is a solution of~\eqref{Equ:GenHomSol}. Hence in any case there is an $h\in\BRSEE\setminus\{0\}$ with~\eqref{Equ:GenHomSol}.\\
We can reorder $\dfield{\BRSEE}{\sigma}$ to
$\BRSEE=\HH[s_1]\dots[s_e]$ with
$\HH=\GG\lr{\tilde{t}_1}\dots\lr{\tilde{t}_{\tilde{u}}}\lr{t_1}\dots\lr{t_u}$
where
$(s_1,\dots,s_{v+\tilde{v}})=(\tilde{\tau}_1,\dots,\tilde{\tau}_{\tilde{v}},\tau_1,\dots,\tau_v)$.
Set $\BRSEE'_j=\HH[s_1]\dots[s_{j}]$. Suppose there is no such $h$
with $h\in\HH\setminus\{0\}$. Then we can choose among all the
possible solutions $h$ with~\eqref{Equ:GenHomSol} an element
$h\in\BRSEE'_k\setminus\BRSEE'_{k-1}$ with $k>0$ being minimal. We
can write $h=\alpha\,s_k^{\mu}+\beta$ with $\mu>0$ and
$\alpha,\beta\in\BRSEE'_{k-1}$ where $\alpha\neq0$.
Doing coefficient comparison w.r.t.\ $s_k^{\mu}$ in~\eqref{Equ:GenHomSol}, using $\sigma(s_k)-s_k\in\BRSEE'_{k-1}$ and knowing that $a_0,\dots,a_s\in\GG$, we conclude that $\alpha$ is a solution of~\eqref{Equ:GenHomSol}; a contradiction to the minimality of $k$.\\
Summarizing, we can find $h\in\HH\setminus\{0\}$ with~\eqref{Equ:GenHomSol}. Now write $$h=\sum_{(\tilde{l}_1,\dots,\tilde{l}_{\tilde{u}},l_1,\dots,l_u)\in S}h_{(\tilde{l}_1,\dots,\tilde{l}_{\tilde{u}},l_1,\dots,l_u)}\tilde{t}_1^{\tilde{l}_1}\dots \tilde{t}_{\tilde{u}}^{\tilde{l}_u}t_1^{l_1}\dots t_u^{l_u}$$
for a finite set $S\subseteq\ZZ^{\tilde{u}+u}$ and $h_{(\tilde{l}_1,\dots,l_u)}\in\GG$.
Since $h\neq0$, we can take $w=h_{(\tilde{l}_1,\dots,l_u)}\in\GG^*$ for some $(\tilde{l}_1,\dots,l_u)\in S$.\\ By coefficient comparison it follows that $h'=w\,t_1^{\tilde{l}_1}\dots t_{u}^{l_u}\neq0$ is a solution of~\eqref{Equ:GenHomSol}.\\
(2) Let $f\in\AR\setminus\{0\}$ and $g\in\AR\lr{t_1}\dots\lr{t_u}$ with~\eqref{Equ:GenInhomSol}. Note that we can write $g=\sum_{(l_1,\dots,l_u)\in S}h_{(l_1,\dots,l_u)}t_1^{l_1}\dots t_u^{l_u}$ for a finite set $S\subseteq\ZZ^u$ and $h_{(l_1,\dots,l_u)}\in\AR$. By coefficient comparison w.r.t.\ $t_1^{0}\dots t_u^{0}$ in~\eqref{Equ:GenInhomSol} we conclude that for $h=h_{(0,\dots,0)}\in\AR\setminus\{0\}$ the equation $a_s\,\sigma^s(h)+\dots+a_0\,h=f$ holds.\qed
\end{proof}

\noindent We will reformulate Proposition~\ref{Prop:ConstraintForRecSol} for homogeneous difference equations to Corollary~\ref{Cor:ParticularImpliesHom} by using the following lemma; for a simpler version see~\cite[Prop.~6.13]{Schneider:05c}.

\begin{mlemma}\label{Lemma:MergeToRPi}
Let $\dfield{\BRSEE}{\sigma}$ with $\BRSEE=\AR\lr{t_1}\dots\lr{t_e}$
be a simple \rpiE-extension of $\dfield{\AR}{\sigma}$ with
$\alpha_i:=\frac{\sigma(t_i)}{t_i}\in\AR^*$, and let $l_i\in\ZZ$
such that $t^{l_1}\dots t^{l_e}\notin\AR$. Then there exists an
\rpiE-extension $\dfield{\AR\lr{t}}{\sigma}$ of
$\dfield{\AR}{\sigma}$ with $\sigma(t)=\alpha\,t$ where
$\alpha=\prod_{i=1}^e\alpha_i^{l_i}$.
\end{mlemma}
\begin{proof}
Define $M=\{i\mid l_i\neq0\}$ and set $h:=t^{l_1}\dots t^{l_e}$.
Note that $\sigma(h)=\alpha\,h$. First, suppose that $\alpha^n=1$
for some $n>0$. Then for all $i\in M$, $t_i$ is an \rE-monomial. In
particular, since the $\alpha_i$ are roots of unity, $\alpha$ is a
root of unity. Let $m>0$ be minimal such that $\alpha^m$. If $m=1$,
then $\alpha=1$ thus $\sigma(h)=h$, and consequently
$h\in\const{\AR}{\sigma}$. Therefore $h=1$, a contradiction. Thus
$\alpha$ is a primitive $m$th root of unity with $m>1$. Now
construct the $A$-extension $\dfield{\AR[t]}{\sigma}$ of
$\dfield{\AR}{\sigma}$ with $\sigma(t)=\alpha\,t$ and suppose that
there is a $k\in\NN$ with $k<m$ and $g\in\AR\setminus\{0\}$ such
that $\sigma(g)=\alpha^k\,g$. We can find an $r$ such that
$t_r^{l_r\,k}\neq1$ (otherwise $h^k=1$, thus
$1=\sigma(h^k)=\alpha^k\,h^k=\alpha^k$ and thus $m$ is not minimal
with $\alpha^m=1$). But this implies that $\alpha_r^{l_r\,k}\neq1$
(otherwise $t_r^{l_r\,k}\in\const{\BRSEE}{\sigma}\setminus\AR$, but
$\const{\BRSEE}{\sigma}=\const{\AR}{\sigma}$). Choose $r\geq1$ to be
maximal with this property, and let $u>1$ be minimal such that
$\alpha_r^u=1$. Then we can find $k'$ with $1\leq k'<u$ with
$\alpha_r^{l_r\,k}=\alpha_r^{k'}$. Further, with
$\tilde{h}=g/(t_1^{l_1}\dots
t_{r-1}^{l_{r-1}})^k\in\AR\lr{t_1}\dots\lr{t_{r-1}}$ we get
$\sigma(\tilde{h})=\alpha_r^{l_r\,k}\,\tilde{h}$.  Hence $t_r$ is not an \rE-monomial by part~3 of Theorem~\ref{Thm:RPSCharacterization}; a contradiction.\\
Otherwise, suppose that there is no $n>0$ with $\alpha^n=1$. Then there is at least one $i\in M$ such that $t_i$ is a \piE-monomial. W.l.o.g.\ suppose that $t_r$ is a \piE-monomial with $\max(M)=r$; otherwise we reorder the generators accordingly. Suppose that the $P$-extension $\dfield{\AR\lr{t}}{\sigma}$ of $\dfield{\AR}{\sigma}$ with $\sigma(t)=\alpha\,t$ is not a \piE-extension. Then there is a $k\in\ZZ\setminus\{0\}$ and $g\in\AR\lr{t_1}\dots\lr{t_{r-1}}\setminus\{0\}$ with $\sigma(g)=\alpha^k\,g$. Define $\tilde{h}=g/(t_1^{l_1}\dots t_{r-1}^{l_{r-1}})^k$. Then, as above, $\sigma(\tilde{h})=\alpha_r^{l_{r}\,k}\,\tilde{h}$ with $l_{r}\,k\neq0$ and consequently $t_r$ is not a \piE-monomial by part~(2) of Theorem~\ref{Thm:RPSCharacterization}, a contradiction.\qed
\end{proof}

\begin{mcorollary}\label{Cor:ParticularImpliesHom}
Let $\dfield{\AR}{\sigma}$ be a simple \rpisiE-extension of
$\dfield{\GG}{\sigma}$ with $a_0,\dots,a_s\in\GG$. If there is a
$\GG$-simple \rpisiE-extension $\dfield{\BRSEE}{\sigma}$ of
$\dfield{\AR}{\sigma}$ in which one finds an
$h\in\BRSEE\setminus\AR$ with~\eqref{Equ:GenHomSol}, then there is
an \rpiE-extension $\dfield{\AR\lr{t}}{\sigma}$ of
$\dfield{\AR}{\sigma}$ with $\frac{\sigma(t)}{t}\in\GG^*$ in which
one finds a solution of~\eqref{Equ:GenHomSol} with $h=w\,t^m$ where
$m\in\ZZ\setminus\{0\}$ and $w\in\GG\setminus\{0\}$.
\end{mcorollary}
\begin{proof}
Suppose there is a $\GG$-simple \rpisiE-extension
$\dfield{\BRSEE}{\sigma}$ of $\dfield{\AR}{\sigma}$ as in
Proposition~\ref{Prop:ConstraintForRecSol} in which we find an
$h\in\BRSEE\setminus\{0\}$ with~\eqref{Equ:GenHomSol}. Then by
part~1 of Proposition~\ref{Prop:ConstraintForRecSol} we can find an
$h'=w\,{\tilde{t}_1}^{\tilde{l}_1}\dots
{\tilde{t}_u}^{\tilde{l}_u}\,t_1^{l_1}\dots t_u^{l_u}\notin\AR$ with
$\tilde{l}_i,l_i\in\ZZ$ and $w\in\GG\setminus\{0\}$
with~\eqref{Equ:GenHomSol} (where $h$ is replaced by $h'$). Set
$a={\tilde{t}_1}^{\tilde{l}_1}\dots
{\tilde{t}_u}^{\tilde{l}_u}\,t_1^{l_1}\dots t_u^{l_u}$ and define
$\alpha:=\frac{\sigma(a)}{a}\in\GG^*$. Then we can construct the
\rpiE-extension $\dfield{\AR\lr{t}}{\sigma}$ of
$\dfield{\AR}{\sigma}$ with $\sigma(t)=\alpha\,t$ by
Lemma~\ref{Lemma:MergeToRPi}. By construction it follows that
$h''=w\,t$ is also a solution of~\eqref{Equ:GenHomSol}.\qed
\end{proof}

With these new results in \rpisiE-theory, we can continue to tackle Problem~C. Recall that we assume that there is no $h\in\AR$ with~\eqref{Equ:DEHomConst}, but there exists a $\GG$-simple \rpisiE-extension in which we find such an $h$. Then by Corollary~\ref{Cor:ParticularImpliesHom} there is also an \rpiE-extension $\dfield{\AR\lr{t}}{\sigma}$ of $\dfield{\AR}{\sigma}$ with $\frac{\sigma(t)}{t}\in\GG^*$ and
$h\in\AR\lr{t}$ with $h=w\,t^m$ where $m\in\ZZ\setminus\{0\}$ and $w\in\GG\setminus\{0\}$ such that~\eqref{Equ:DEHomConst} holds. As above, we can consider two cases.

\medskip

\noindent\textbf{Case~2.1:} Suppose that we find a $\gamma\in\AR\lr{t}$ with~\eqref{Equ:DEInhomConst}. Then with Lemma~\ref{Lemma:FindConstantsInside} we get $g=g_0\,x_0+\dots+g_s\,x_s+g_{s+1}$ with $g_i\in\AR\lr{t}$ such that $\sigma(g)=g$ holds. Even more, looking at the construction it follows for $0\leq i\leq s$ that $g_i=g'_i\,t^m$ for some $g'_i\in\GG$. Further we can use the following simple lemma.

\begin{mlemma}
Let $\dfield{\AR\lr{t}}{\sigma}$ be an \rpiE-extension of $\dfield{\AR}{\sigma}$. Let $f=f'\,t^m$ with $f'\in\AR$, $m\neq0$, and $g\in\AR\lr{t}$. If $\sigma(g)-g=f$, then $g=g'\,t^m+c$ with $g'\in\AR$ and $c\in\const{\AR}{\sigma}$.
\end{mlemma}
\begin{proof}
Let $\alpha=\frac{\sigma(t)}{t}\in\AR^*$ and $g=\sum_i g_i\,t^i\in\AR\lr{t}$. By coefficient comparison it follows that $\alpha^i\,\sigma(g_i)-g_i=0$ for all $i$ with $i\neq m$. By part~(2) of Theorem~\ref{Thm:RPSCharacterization} it follows that $g_i=0$ if $i\neq0$. Further, $g_i\in\const{\AR}{\sigma}$ if $i=0$. \qed
\end{proof}

\noindent Applying this lemma to~\eqref{Equ:DEInhomConst}, we conclude that we can choose
$g_{s+1}=g'_{s+1}\,t^m$ for some $g'_{s+1}\in\AR$ and therefore
\begin{equation}\label{Equ:ContgInRP}
g=t^m(g'_0\,x_0+\dots+g'_s\,x_s+g'_{s+1})
\end{equation}
with $g'_i\in\GG$ for $0\leq i\leq s$, $g'_{s+1}\in\AR$ and $m\in\ZZ\setminus\{0\}$.

\medskip

\noindent\textbf{Case 2.2:} There is no $\gamma\in\AR\lr{t}$. Then as in Case~1.2
we can construct the \sigmaE-extension $\dfield{\AR\lr{t}[\tau]}{\sigma}$ of $\dfield{\AR\lr{t}}{\sigma}$ with $\sigma(\tau)=\tau-a_{s+1}\,\sigma(h)$ and get
\begin{equation}\label{Equ:ContgInRPS}
g=t^m(g'_0\,x_0+\dots+g'_s\,x_s)+\tau
\end{equation}
with $m\in\ZZ\setminus\{0\}$ and $g_i\in\GG$ for $0\leq i\leq s$.

\medskip

\noindent Summarizing, we obtain the following result.

\begin{mtheorem}\label{Thm:ConstantChar}
Let $\dfield{\AR}{\sigma}$ be a simple \rpisiE-extension of
$\dfield{\GG}{\sigma}$ and let $\dfield{\HH}{\sigma}$ be a h.o.l.\
extension of $\dfield{\AR}{\sigma}$ with $\HH=\AR[x_0,\dots,x_s]$
and~\eqref{HOEShift} where $a_0,\dots,a_a\in\GG$ and
$a_{s+1}\in\AR$. Suppose that there is an $h$ in a $\GG$-simple
\rpisiE-extension with~\eqref{Equ:DEHomConst}. Then there is a
$\GG$-simple \rpisiE-extension $\dfield{\BRSEE}{\sigma}$ of
$\dfield{\AR}{\sigma}$ and a h.o.l.\ extension
$\dfield{\BRSEE[x_1,\dots,x_s]}{\sigma}$ of
$\dfield{\BRSEE}{\sigma}$ with~\eqref{HOEShift} in which one gets
$g\in\BRSEE[x_1,\dots,x_s]$ with $\sigma(g)=g$. In particular, one
of the four situations hold.
\begin{description}
 \item[Case 1.1:] \eqref{Equ:ConstgInA} with $g_i\in\AR$.
 \item[Case 1.2:] $\dfield{\BRSEE}{\sigma}$ is a \sigmaE-extension of $\dfield{\AR}{\sigma}$ with $\BRSEE=\AR[\tau]$; \eqref{Equ:ConstgInSigma} with $g_i\in\AR$.
 \item[Case 2.1:] $\dfield{\BRSEE}{\sigma}$ is an \rpiE-extension of $\dfield{\AR}{\sigma}$ with $\BRSEE=\AR\lr{t}$ and $\frac{\sigma(t)}{t}\in\GG^*$; \eqref{Equ:ContgInRP} with $g'_i\in\GG$ for $0\leq i\leq s$ and $g'_{s+1}\in\AR$.
 \item[Case 2.2:] $\dfield{\BRSEE}{\sigma}$ is an \rpisiE-extension $\dfield{\AR}{\sigma}$ with $\BRSEE=\AR\lr{t}[\tau]$ where $\frac{\sigma(t)}t\in\GG$ and $\sigma(\tau)-\tau\in\AR$; \eqref{Equ:ContgInRPS} with $g'_i\in\GG$ for $0\leq i\leq s$.
 \end{description}
\end{mtheorem}

\noindent Let $\dfield{\AR}{\sigma}$ be a simple \rpisiE-extension of a \pisiE-field $\dfield{\GG}{\sigma}$ and suppose that $a_0,\dots,a_s\in\GG$ and $a_{s+1}\in\AR$. Then we can tackle problem~C as follows.
\begin{enumerate}
 \item Decide constructively if there is an $h\in\AR\setminus\{0\}$ with~\eqref{Equ:DEHomConst} using the algorithms from~\cite{Bron:00,Schneider:01,Schneider:05b,ABPS:17,DR1,DR2}. If such an $h$ exists, continue with step~6.

 \item Otherwise, decide constructively if there is an \rpiE-extension $\dfield{\GG\lr{t}}{\sigma}$ of $\dfield{\GG}{\sigma}$ such that we find $h\in\GG\lr{t}$ with~\eqref{Equ:DEHomConst}. Here one can utilize, e.g., the algorithms given in~\cite{petkovsek1996b,BAUER1999711} if $\dfield{\GG}{\sigma}$ is one of the instances (1--3) from Example~\ref{Exp:BaseFields}. Otherwise, we can utilize the more general algorithms from~\cite{ABPS:17}.

 \item Check if there is an $h'\in\AR$ with $\frac{\sigma(h')}{h'}=\frac{\sigma(t)}{t}$ using the algorithms from~\cite{DR1}.
 If yes, $h'$ is a solution of~\eqref{Equ:DEHomConst}. Go to step~6 where $h'$ takes over the role of $h$.

 \item Check if the $AP$-extension $\dfield{\AR\lr{t}}{\sigma}$ of $\dfield{\AR}{\sigma}$ is an \rpiE-extension using Theorem~\ref{Thm:RPSCharacterization} and applying the algorithms from~\cite{DR1}. If yes, we get the solution $h'=t\in\AR\lr{t}$ of~\eqref{Equ:DEHomConst} and we go to step~6 where $h'$ takes over the role of $h$.

 \item Try to redesign and extend the difference ring $\dfield{\AR}{\sigma}$ to $\dfield{\AR'}{\sigma}$ such that one can find $h\in\AR'$ with~\eqref{Equ:DEHomConst} and such that $\dfield{\AR'}{\sigma}$ is an \rpisiE-extension of $\dfield{\GG}{\sigma}$. If $\dfield{\GG}{\sigma}$ is one of the instances (1--3) from Example~\ref{Exp:BaseFields}, this can be accomplished with the algorithms from~\cite{Schneider:05c,OS:17} in combination with~\cite{DR1}. For a general \pisiE-field $\dfield{\GG}{\sigma}$ our method might fail. Otherwise replace $\AR$ by $\AR'$ and go to step~6.

 \item Compute, if possible, a $\gamma\in\AR$ with~\eqref{Equ:DEInhomConst} using the algorithms from~\cite{DR1}. If this is not possible, construct the \sigmaE-extension $\dfield{\AR[\tau]}{\sigma}$ of $\dfield{\AR}{\sigma}$ with $\sigma(\tau)=\tau-a_{s+1}\,\sigma(h)$ and set $\gamma=\tau$.

 \item Use Lemma~\ref{Lemma:FindConstantsInside} with the given $h,\gamma$ to compute $g$ with $\sigma(g)=g$.
 \end{enumerate}

\begin{mremark}\label{Remark:AlgConstant}
(1) If $\dfield{\GG}{\sigma}$ is one of the base difference fields (1--3) from Example~\ref{Exp:BaseFields}, all steps can be carried out. However, if $\dfield{\GG}{\sigma}$ is a general \pisiE-field,
one might fail in step~5 with the existing algorithms to redesign and extend the difference ring $\dfield{\AR}{\sigma}$ to $\dfield{\AR'}{\sigma}$ such that it is an \rpisiE-extension of $\dfield{\GG}{\sigma}$ in which one gets $h\in\AR'$ with~\eqref{Equ:DEHomConst}.\\
(2) Suppose that there exists a $g=g_0\,x_0+\dots+g_s+g_{s+1}$ with
$g_i\in\BRSEE$ for $0\leq i\leq s+1$ and $\sigma(g)=g$ for some
$\GG$-simple \rpisiE-extension $\dfield{\BRSEE}{\sigma}$ of
$\dfield{\AR}{\sigma}$. Then the above method will always find such
a $g$ as predicted in Theorem~\ref{Thm:ConstantChar}. Namely, by
Lemma~\ref{Lemma:FindConstantsInside} there is an $h\in\BRSEE$
with~\eqref{Equ:DEHomConst}. Hence we may either assume that there
is a solution of~\eqref{Equ:DEHomConst} in $\AR$ or by
Corollary~\ref{Cor:ParticularImpliesHom} there is an \rpiE-extension
$\dfield{\AR\lr{t}}{\sigma}$ of $\dfield{\AR}{\sigma}$ with
$\frac{\sigma(t)}{t}\in\GG^*$ in which we can find a solution
of~\eqref{Equ:DEHomConst} in $\GG\lr{t}$. Thus the above method can
be executed without entering in step~5.
\end{mremark}

\begin{mexample}\label{Exp:ConstantReduction}
Consider the sequence~\eqref{Equ:XOrder2Triv} with the initial values $X(0)=0$, $X(1)=-1$. We remark that this recurrence is completely solvable in terms of d'Alembertian solutions:
\begin{equation}\label{Equ:SimpleClosedForm}
X(k)=2^{k}(S_1(\tfrac12,k)
        -
        S_1(k)).
\end{equation}
We will compute this zero-order recurrence by iteratively computing constants. We start with~\eqref{Equ:XOrder2Triv} and set up the underlying h.o.l.\ extension. Since there is a recurrence with smaller order (order zero), there must exist a non-trivial constant. Our algorithm produces the constant
$G(k)=2^{-k}
+(1+k) 2^{1-k} X(k)
+(-1-k) 2^{-k} X(1+k).
$
With $G(0)=2$ we get a new recurrence of order $1$:
$X(1+k)=\frac{-1+2^{1+k}}{-1-k}
+2 X(k).$
We use this recurrence and set up a new h.o.l.\ extension and search again for a constant. We get
$G(k)=2^{-k} X(k)-S_1(\tfrac12,k)
        +
        S_1(k)$
and with $G(0)=0$ we obtain~\eqref{Equ:SimpleClosedForm}.
\end{mexample}

In other words, computing stepwise constants (where in each step the constant has the shape as worked out in Theorem~\ref{Thm:ConstantChar}), we find the smallest possible recurrence that can be given in terms of simple \rpisiE-extensions. Note that this mechanism has been utilized already earlier to find an optimal recurrence in the context of finite element methods~\cite{BPPRSS:06}. In particular, if there is a recurrence of order $0$ where the inhomogeneous part is given in a simple \rpisiE-extension, such a recurrence will be eventually calculated with our method from above. Note that this strategy to find minimal recurrences (and to solve the recurrence in terms of d'Alembertian solutions if possible) is also related to the remarks given in~\cite[page~163]{petkovsek1996b} that deals with the computation of left factors of a recurrence.

\subsubsection{The refined parameterized telescoping problem}\label{Subsec:RPTHolo}

Suppose that we are given $F_i(k)$ for $1\leq i\leq d$ and suppose that we can represent them in a difference ring as introduced above. Namely, suppose that we succeeded in constructing an \rpisiE-extension $\dfield{\AR}{\sigma}$ of a \pisiE-field $\dfield{\GG}{\sigma}$ with $\KK=\const{\GG}{\sigma}$, and on top of this, we designed a h.o.l.\ extension $\dfield{\AR[x_1,\dots,x_s]}{\sigma}$ with~\eqref{HOEShift} where $a_0,\dots,a_s\in\GG$ and $a_{s+1}\in\AR$ with the following property: the $F_i(k)$ can be rephrased as
\begin{equation}\label{Equ:fiInX}
f_i=f_{i,0}\,x_0+\dots+f_{i,s}\,x_s+f_{i,s+1}
\end{equation}
for $1\leq i\leq d$ with $f_{i,j}\in\GG$ for $0\leq j\leq s$ and $f_{i,s+1}\in\AR$.

In this setting, we are interested in solving Problem~RPT. Namely,
we aim at finding, if possible, an appropriate $\GG$-simple
\rpisiE-extension $\dfield{\BRSEE}{\sigma}$ of
$\dfield{\AR}{\sigma}$ in which one can solve Problem~RPT with
$c_1,\dots,c_d\in\KK$ where $c_1\neq0$ and
$g=g_0\,x_0+\dots+g_s\,x_s+g_{s+1}$ with $g_i\in\BRSEE$ for $0\leq
i\leq s+1$. Then given such a result and rephrasing $g$ as $G(k)$
yields~\eqref{eq:ptcert} and enables one to compute the sum
relation~\eqref{eq:ptproblem}.

In our main application we set $F_i(k)=F(n+i-1,k)$ for a bivariate sequence. Then~\eqref{eq:ptproblem} can be turned to a linear recurrence of the form~\eqref{Equ:Rec} for a definite sum, say $S(n)=\sum_{k=l}^{L(n)}F(n,k)$ for $l\in\NN$ and for some integer linear function $L(n)$. During this construction, we should keep in mind that various optimality criteria might lead to different preferable recurrences.
\begin{enumerate}
 \item Find a recurrence~\eqref{Equ:Rec} with lowest order $d-1$.
 \item Find a recurrence such that the underlying difference ring is as simple as possible (e.g., the number of arising sums and products is as small as possible, the nesting depth of the sums is minimal, or the number of objects within the summands is as low as possible.)
\end{enumerate}

Note that in most examples both criteria cannot be fulfilled simultaneously: in an appropriate \rpisiE-extension the number $d$ might be reduced, but the difference ring will be enlarged by further, most probably more complicated \rpisiE-monomials; in the extreme case one might find a zero-order recurrence formulated in a rather large \rpisiE-extension.
Contrary, increasing $d$ might lead to simpler \rpisiE-extensions in which the recurrence can be formulated; ideally, one can even find a recurrence without introducing any further \rpisiE-monomials.
In our experience a compromise between these extremes are preferable to reduce the underlying calculation time. On one side, we are interested in calculating a recurrence as efficiently as possible. On the other side, we might use the found recurrence as the new defining h.o.l.\ extension and to tackle another parameterized telescoping problem in a recursive fashion (see Section~\ref{Sec:MultiSumApproach}). Hence the derivation of a good recurrence (not to large in $d$ but also not too complicated objects in the inhomogeneous part) will be an important criterion.

Having this in mind, we will focus now on various tactics to tackle Problem~RPT that give us reasonable flexibility for tackling definite multi-sums but will be not too involved concerning the complexity of the underlying algorithms.
We start as follows.
Set $f=c_1\,f_1+\dots+c_d\,f_d$ for unknown $c_1,\dots,c_d$ and write $f=h_0\,x_0+\dots+h_s\,x_s+h_{s+1}$ with $h_i=c_1\,f_{i,1}+\dots+c_d\,f_{i,d}$. By Lemma~\ref{Lemma:IndefiniteConstraint} it follows that~\eqref{Equ:HigherRec} and~\eqref{Equ:InHomTele} must hold (where $f_i$ is replaced by $h_i$). Note that~\eqref{Equ:HigherRec}
reads as
\begin{equation}\label{Equ:TildeParaEqu}
\sum_{j=0}^s\sigma^{s-j}(a_j)\sigma^{s-j+1}(g_s)-g_s=\sum_{j=0}^s\sigma^{s-j}(h_{j})=c_1\,\tilde{f_1}+\dots+c_d\,\tilde{f_d}
\end{equation}
with
\begin{equation}\label{Equ:ComputefTilde}
\tilde{f}_i=\sum_{j=0}^s\sigma^{s-j}(f_{i,j})\in\GG.
\end{equation}

\noindent Hence one could utilize the summation package \texttt{Sigma} as follows: (1) look for a $g_s$ in $\GG$; (2) if this fails, try to find a solution in $\AR$; (3) if there is no such solution, search for a solution in a $\GG$-simple \rpisiE-extension.

\begin{mremark}\label{Remark:ExplainRecCoeff}
Assume there exists such a solution $g_s\neq0$
with~\eqref{Equ:TildeParaEqu} either in $\dfield{\AR}{\sigma}$ or in
a $\GG$-simple \rpisiE-extension $\dfield{\BRSEE}{\sigma}$ of
$\dfield{\GG}{\sigma}$, but not in $\GG$. This implies that there is
an $h\in\BRSEE$ with~\eqref{Equ:DEHomConst}: if the right hand side
of~\eqref{Equ:TildeParaEqu} is $0$, we can set $h:=g_s$. Otherwise,
we utilize Proposition~\ref{Prop:ConstraintForRecSol} (by taking the
special case $\AR=\GG$). Namely, part~2 implies that a solution
of~\eqref{Equ:TildeParaEqu} must depend on a \sigmaE-monomial that
is introduced by the extension $\dfield{\BRSEE}{\sigma}$ of
$\dfield{\AR}{\sigma}$. Finally, part~1 implies that there is an
$h\in\BRSEE$  with~\eqref{Equ:DEHomConst}. Hence we can utilize
Theorem~\ref{Thm:ConstantChar} and it follows that we can construct
a $\GG$-simple \rpisiE-extension in which one can compute
$g'=g'_0\,x_0+\dots+g'_s\,x_s+g'_{s+1}$ with $\sigma(g')=g'$ and
$g_s\neq0$. Note that such an extension and $g'$ can be even
computed; see part~(2) of Remark~\ref{Remark:AlgConstant}. Hence
following the recipe after Lemma~\ref{Lemma:FindConstantsInside} we
can construct a recurrence~\eqref{Equ:ShorterXRec} ($g'_i$ is
rephrased as the summation object $G_i(k)$) for our summation object
$X(k)$ which has a smaller recurrence order. Further, we can
construct an improved h.o.l.\ extension with recurrence order $s$
that describes better the shift behavior of the sequence $X(k)$.
\end{mremark}

\noindent Summarizing, finding a solution $g_s$ in $\AR$ or in a $\GG$-simple \rpisiE-extension implies that one can also reduce the recurrence order of the h.o.l.\ extension that models $X(k)$.
In this regard, we emphasize that having a recurrence with smaller order will also increase the chance to find a parameterized telescoping solution:
the larger the recurrence order is, the more sequences are satisfied by the recurrence and thus a solution of Problem~PRS is more general. Conversely, the smaller the recurrence order is, the better the problem description and thus the higher the chances are to find a solution if it exists.
Hence instead of searching for a $g_s$ in $\AR$ or in an appropriate $\GG$-simple \rpisiE-extension, we opt for outsourcing this task to the user: if it seems appropriate, the user should try to hunt for a recurrence with lower order by either using other summation tactics (see Example~\ref{Exp:PiSi}) or applying the machinery mentioned in part~(2) of Remark~\ref{Remark:AlgConstant} as a preprocessing step to produce a h.o.l.\ extension with lower order.

\begin{mremark}
 Note that the classical holonomic summation algorithms~\cite{CHYZAK2000115,Koutschan:10} handle the case~\eqref{HOEShift} with $a_{s+1}=0$ and $a_i\in\GG$ where $\dfield{\GG}{\sigma}$ is the rational or $q$-rational difference field (see instances~(1) and~(2) of Example~\ref{Exp:BaseFields}). In most cases, the arising recurrences are optimal in the following sense: the recurrence orders cannot be reduced and the recurrence system is the defining relation. Together with the above inside (see Remark~\ref{Remark:ExplainRecCoeff}) this explains why standard holonomic approaches are optimal: they hunt for solutions $g=g_0\,x_0+\dots+g_s\,x_s$ where the $g_i$ are in $\GG$ and do not try to look for any simple \rpisiE-extension.
\end{mremark}

With this understanding, we will restrict\footnote{If one is only interested in the telescoping problem with $d=1$, it might be worthwhile to look for a solution of~\eqref{Equ:HigherRec} in a $\GG$-simple \rpisiE-extension; this particular case is neglected in the following.} ourselves to the following

\medskip

\noindent\textbf{Strategy 1:} we will search for $c_1,\dots,c_d\in\KK$ with $c_1\neq0$ and for $g_s$ with~\eqref{Equ:TildeParaEqu} and~\eqref{Equ:ComputefTilde} only in $\GG$, but not in $\AR$ or in any other simple \rpisiE-extension. To obtain all such solutions, we will assume that we can solve the following subproblem.

\begin{center}
\noindent\fbox{\begin{minipage}{11.4cm}
\noindent \textbf{Problem PRS in $\dfield{\GG}{\sigma}$:} Parameterized recurrence solving.\\
\small
\textit{Given} a difference field $\dfield{\GG}{\sigma}$ with constant field $\KK=\const{\GG}{\sigma}$, $\vect{0}\neq(\tilde{a}_0,\dots,\tilde{a}_m)\in\GG^{m+1}$ and $\vect{\tilde{f}}=(\tilde{f}_1,\dots,\tilde{f}_d)\in\GG^d$.
\textit{Find} a basis of the $\KK$-vector space\footnote{The dimension of $V$ is at most $d+m$; see~\cite{Cohn:65,Schneider:05b}.}
\begin{equation}\label{Equ:ParaRecSol}
V=\{(c_1,\dots,c_d,g)\in\KK^d\times\GG\mid \tilde{a}_0\,g+\dots+\tilde{a}_m\,\sigma^m(g)=c_1\,\tilde{f}_1+\dots+c_d\,\tilde{f}_d\}.
\end{equation}
\end{minipage}}
\end{center}

\noindent We remark that this strategy is also down to earth: searching $g_s$ in $\GG$ is usually very efficient and does not need any fancy algorithms. In particular, we can solve Problem~PRS if $\dfield{\GG}{\sigma}$ is a \pisiE-field\footnote{For the rational and $q$-rational difference fields see also~\cite{Abramov:89a,Abramov1995,Hoeij:98}}; see~\cite{Bron:00,Schneider:01,Schneider:05b,ABPS:17}.



\medskip

We continue with our algorithm for Problem~RPT. Namely, suppose that we can compute a non-empty basis of $V$ as posed in Problem~PRS. Then by Lemma~\ref{Lemma:IndefiniteConstraint} we have to find a $(c_1,\dots,c_d,g_s)\in V$ with $c_1\neq0$ such that
there is a $g_{s+1}$ with~\eqref{Equ:InHomTele} where $f_{s+1}$ must be replaced by $c_1\,f_{1,s+1}+\dots+c_d\,f_{d,s+1}$. If we find such a $g_{s+1}$ in $\AR$, we are done. Namely, following Lemma~\ref{Lemma:IndefiniteConstraint} we take
\begin{equation}\label{Equ:giPara}
\begin{split}
g_0&=a_0\,g_s-(c_1\,f_{0,1}+\dots+c_d\,f_{0,d}),\\
g_i&=\sigma(g_{i-1})+a_i\,\sigma(g_s)-(c_1\,f_{i,1}+\dots+c_d\,f_{i,d}),\quad 1\leq i<s
\end{split}
\end{equation}
and get the desired solution $g=g_0\,x_0+\dots+g_s\,x_s+g_{s+1}$ of~\eqref{Equ:ParaDF}.
Otherwise, one could take any  $(c_1,\dots,c_d,g_s)\in V$ with $c_1\neq0$. Then by Theorem~\ref{Thm:RPSCharacterization} we can construct the \sigmaE-extension $\dfield{\AR[\tau]}{\sigma}$ of $\dfield{\AR}{\sigma}$ with $\sigma(\tau)=\tau+\phi$ where
\begin{equation}\label{Equ:BetaPara}
\phi=(c_1\,f_{1,s+1}+\dots+c_d\,f_{d,s+1})-a_{s+1}\,\sigma(g_{s})
\end{equation}
and can choose $g_{s+1}=\tau$.
However, this might not be the best choice. Using strategies $\text{RPT}_r$ with $r=2,3,4$ of Remark~\ref{Remark:RPTVariants} might produce a better result. In any case, if one fails to find a solution with the proposed tactics or if the produced \rpisiE-extension is too involved for further processing (in particular for our application in Section~\ref{Sec:MultiSumApproach}), one can also enlarge $d$ to search for a recurrence with a higher order but with simpler summation objects involved. These considerations yield

\medskip

\noindent\textbf{Strategy 2:} we will search $g_{s+1}$ in $\AR$
($\text{RPT}_1$) and if this fails provide the option to use our
refined algorithms $\text{RPT}_2$, $\text{RPT}_3$ or $\text{RPT}_4$
of Remark~\ref{Remark:RPTVariants}  to look for an optimal
\sigmaE-extension $\dfield{\BRSEE}{\sigma}$ of
$\dfield{\AR}{\sigma}$ in which $g_{s+1}$ can be found.

\medskip

Summarizing, we propose the following general summation tactic in higher-order extensions that enables one to
incorporate our Strategies~1 and~2 from above.

\begin{alg}{\label{AlgParaTele}}{Refined holonomic parameterized telescoping.}
{\algname{ParameterizedTelescoping}($\dfield{\AR[x_0,\dots,x_s]}{\sigma},\vect{f}$)}
{A difference ring extension $\dfield{\AR}{\sigma}$ of a difference field $\dfield{\GG}{\sigma}$ with constant field $\KK=\const{\GG}{\sigma}=\const{\AR}{\sigma}$ where one can solve Problems~RPT in $\dfield{\AR}{\sigma}$ and PRS in $\dfield{\GG}{\sigma}$. A h.o.l.\ extension $\dfield{\AR[x_0,\dots,x_s]}{\sigma}$ of
$\dfield{\AR}{\sigma}$ with~\eqref{HOEShift} where $a_0,\dots,a_s\in\GG$ and $a_{s+1}\in\AR$.
$\vect{f}=(f_1,\dots,f_d)$ with~\eqref{Equ:fiInX} for $1\leq i\leq d$ where $f_{i,j}\in\GG$ with $1\leq j\leq s$ and $f_{i,s+1}\in\AR$.}
{A h.o.l.\ extension $\dfield{\BRSEE[x_0,\dots,x_s]}{\sigma}$ of
$\dfield{\BRSEE}{\sigma}$ with~\eqref{HOEShift} where
$\dfield{\BRSEE}{\sigma}$ is an ``optimal''\footnotemark extension
of $\dfield{\AR}{\sigma}$ with $g\in \GG\,x_0+\dots+\GG\,x_s+\BRSEE$
and $c_1,\dots,c_d\in\KK$ s.t.\ $c_1\neq0$ and~\eqref{Equ:ParaDF}.
If such an optimal extension does not exist, the output is ``No
solution''.}
\footnotetext{We will make this statement precise in Theorem~\ref{Thm:OptimalDF} by choosing specific variants of Problem~RPT.}
\item Compute $\tilde{f}_i$ for $1\leq i\leq d$ as given in~\eqref{Equ:ComputefTilde}, $\tilde{a}_0=-1$,
$\tilde{a}_i=\sigma^{i-1}(a_{s+1-i})\in\GG$
for $1\leq i\leq s+1$.

\item\label{Alg:SolvePRS} Solve Problem~PRS: compute a basis $B=\{ (c_{i,1},\dots,c_{i,d},\gamma_i)\}_{1\leq i\leq n}$ of~\eqref{Equ:ParaRecSol}.

\item\label{Alg:Quit1} If $B=\{\}$ or $c_{1,1}=c_{1,2}=\dots=c_{1,d}=0$, then Return ``No solution''.

\item\label{Alg:StepSpecialColInC} We assume that $(c_{1,1},\dots,c_{1,n})$ has at most one entry which is non-zero. Otherwise, take one row vector in $B$ where the first entry is non-zero and perform row operations over $\KK$ with the other row vectors of $B$ such that the first entries are zero (note that the result will be again a basis of~\eqref{Equ:ParaRecSol}).

\item Define $\vect{C}=(c_{i,j})\in\KK^{n\times d}$ and $\vect{\gamma}=(\gamma_1,\dots,\gamma_n)\in\GG^n$, and compute
\begin{equation}\label{Equ:PhiCalculate}
\vect{\phi}=(\phi_1,\dots,\phi_n)=\vect{C}\,(f_{1,s+1},\dots,f_{d,s+1})^t-a_{s+1}\,\sigma(\vect{\gamma})^t\in\AR^n.
\end{equation}

\item\label{Alg:SolveRPT} Solve Problem~RPT: find, if possible, an ``optimal'' difference ring extension $\dfield{\BRSEE}{\sigma}$ of $\dfield{\AR}{\sigma}$ with $g_{s+1}\in \BRSEE$ and $\kappa_1,\dots,\kappa_n\in\KK$ with $\kappa_1\neq0$ and $\sigma(g_{s+1})-g_{s+1}=\kappa_1\,\phi_1+\dots+\kappa_{n}\,\phi_n$.

\item\label{Alg:Quit2} If such an optimal extension does not exist, return ``No solution''.

\item\label{Alg:ComputecVec} Otherwise, compute
$(c_1,\dots,c_d)=(\kappa_1,\dots,\kappa_n)\,\vect{C}\in\KK^d$ and $g_s=(\kappa_1,\dots,\kappa_n)\,\vect{\gamma}^t\in\GG$.

\item Compute the $g_i$ with $0\leq i<s$ as given in~\eqref{Equ:giPara}.
\item Return $(c_1,\dots,c_d)\in\KK^d$ and $g=g_0\,x_0+\dots g_s\,x_s+g_{s+1}$.

\end{alg}

\begin{mproposition}\label{Prop:BasicPropOfAlg}
If Algorithm~\ref{AlgParaTele} returns $(c_1,\dots,c_d)\in\KK^d$ and
$g=g_0\,x_0+\dots g_s\,x_s+g_{s+1}$, then $g_1,\dots,g_s\in\GG$,
$g_{s+1}\in\BRSEE$, $c_1\neq0$ and~\eqref{Equ:ParaDF} holds.
\end{mproposition}
\begin{proof}
Suppose that the algorithm outputs $(c_1,\dots,c_d)\in\KK^d$ and
$g=g_0\,x_0+\dots+g_s\,x_s+g_{s+1}$. By construction, $g_i\in\GG$
for $0\leq i\leq s$ and $g_{s+1}\in\BRSEE$. Further, the matrix
$\vect{C}$ has in the first column precisely one nonzero entry (see
step~\eqref{Alg:StepSpecialColInC}). Thus with
$(c_1,\dots,c_d)=(\kappa_1,\dots,\kappa_n)\,\vect{C}$ in
step~\eqref{Alg:ComputecVec} it follows that $c_1\neq0$ if and only
if $\kappa_1\neq0$. But $\kappa_1\neq0$ is guaranteed in
step~\eqref{Alg:SolveRPT} due to the specification of Problem~RPT.
Hence $c_1\neq0$. Finally, define $h=c_1\,f_1+\dots+c_d\,f_d$ and
write $h=h_0\,x_0+\dots+h_s\,x_s+h_{s+1}$ with $h_i\in\AR$ for
$0\leq i\leq s$ and $h_{s+1}\in\BRSEE$. Then
\begin{align*}
\sum_{j=0}^s&\sigma^{s-j}(a_j)\sigma^{s-j+1}(g_s)-g_s=
(\kappa_1,\dots,\kappa_n)\Big(\sum_{j=0}^s\sigma^{s-j}(a_j)\sigma^{s-j+1}(\vect{\gamma}^t)-\vect{\gamma}^t\Big)\\
&=(\kappa_1,\dots,\kappa_n)\,\vect{C}\,(\tilde{f_1},\dots,\tilde{f_d})^t=c_1\,\tilde{f}_1+\dots,+c_d\,\tilde{f}_d\stackrel{\eqref{Equ:ComputefTilde}}{=}\sum_{j=0}^s\sigma^{s-j}(h_{j}),\\
\sigma&(g_{s+1})-g_{s+1}=\kappa_1\,\phi_1+\dots+\kappa_{n}\,\phi_n\\
&=(\kappa_1,\dots,\kappa_n)(\vect{C}(f_{1,s+1},\dots,f_{d,s+1})^t-a_{s+1}\,\sigma(\vect{\gamma})^t)\\
&=(c_1,\dots,c_d)(f_{1,s+1},\dots,f_{d,s+1})^t-a_{s+1}\,(\kappa_1,\dots,\kappa_n)\sigma(\vect{\gamma})^t=h_{s+1}-a_{s+1}\,\sigma(g_s).
\end{align*}
Thus by Lemma~\ref{Lemma:IndefiniteConstraint} it follows that $\sigma(g)-g=f$. \qed
\end{proof}

\begin{mtheorem}\label{Thm:OptimalDF}
Let $\dfield{\AR}{\sigma}$ be a simple \rpisiE-extension of a \pisiE-field $\dfield{\GG}{\sigma}$ with $\AR=\GG\lr{t_{1}}\dots\lr{t_e}$; let $f_1,\dots,f_d\in\AR$ with~\eqref{Equ:fiInX} for $1\leq i\leq d$ where $f_{i,j}\in\GG$ with $1\leq j\leq s$ and $f_{i,s+1}\in\AR$. Execute Algorithm~\ref{AlgParaTele} where in step~\eqref{Alg:SolveRPT} Problem~RPT is specialized by one of the versions $\text{RPT}_r$ with $r=1,2,3,4$. as given in Remark~\ref{Remark:RPTVariants}. If the output is $(c_1,\dots,c_d)\in\KK^d$ and $g=g_0\,x_0+\dots g_s\,x_s+g_{s+1}$, then the following holds for the corresponding specialization.
\begin{description}
\item[$\text{RPT}_1$.] $g_{s+1}\in\AR=\BRSEE$.
\item[$\text{RPT}_2$.] $g_{s+1}$ as given in $\text{RPT}_1$ if this is possible. Otherwise, one gets a \sigmaE-extension $\dfield{\BRSEE}{\sigma}$ of $\dfield{\AR}{\sigma}$ with $g_{s+1}\in\BRSEE\setminus\AR$ and $\depth(g_{s+1})\leq\depth(c_1\,f_{1,s+1}+\dots+c_d\,f_{d,s+1})$.
\item[$\text{RPT}_3$.] $g_{s+1}$ as given in $\text{RPT}_2$ if this is possible. Otherwise, one obtains a \sigmaE-extension $\dfield{\AR[\tau]}{\sigma}$ of $\dfield{\AR}{\sigma}$ with $g_{s+1}\in\AR\setminus\AR$ and $\sigma(\tau)-\tau\in\GG\lr{t_1}\dots\lr{t_i}$ with $0\leq i<e$ where at least one of the $t_{i+1},\dots,t_e$ occurs in $c_1\,f_{1,s+1}+\dots+c_d\,f_{d,s+1}$ and $i$ is minimal
among all such possible solutions.
\item[$\text{RPT}_4$.] $g_{s+1}$ as given in $\text{RPT}_3$ if this is possible. Otherwise, $g_{s+1}=\tau$ within the \sigmaE-extension $\dfield{\AR[\tau]}{\sigma}$ of $\dfield{\AR}{\sigma}$ with $\sigma(\tau)=\tau+\phi$ where~\eqref{Equ:BetaPara}.
\end{description}
If the output is ``No solution'', then there is no solution of~\eqref{Equ:ParaDF} with $(c_1,\dots,c_d)\in\KK^d$ where $c_1\neq0$ and  $g=g_0\,x_0+\dots g_s\,x_s+g_{s+1}$ with $g_i\in\GG$ for $1\leq i\leq s$ and where $g_{s+1}$ can be represented as formulated in $\text{RPT}_r$ with $r=1,2,3,4$, respectively.
\end{mtheorem}
\begin{proof}
Problem~PRS can be solved in a \pisiE-field;
see~\cite{Bron:00,Schneider:01,Schneider:05b,ABPS:17}. Further,
Problems~$\text{RPT}_r$ with $r=1,2,3,4$ can be solved in this
setting; see Remark~\ref{Remark:RPTVariants}. Now let
$r\in\{1,2,3,4\}$ and suppose that our algorithm is executed with
variant $\text{RPT}_r$. If Algorithm~\ref{AlgParaTele} returns
$(c_1,\dots,c_d)\in\KK^d$ and $g=g_0\,x_0+\dots g_s\,x_s+g_{s+1}$,
then $g_1,\dots,g_s\in\GG$, $g_{s+1}\in\BRSEE$, $c_1\neq0$
and~\eqref{Equ:ParaDF} holds by
Proposition~\ref{Prop:BasicPropOfAlg}. Further, by construction the
$g_{s+1}$ is given as specified in $\text{RPT}_r$. This completes
the first part. Now suppose that the algorithm returns ``No
solution'' but there exists a solution $(c_1,\dots,c_d)\in\KK^d$
with $c_1\neq0$ and $g_{s+1}$ as specified in $\text{RPT}_r$. By
Lemma~\ref{Lemma:IndefiniteConstraint} we conclude that
$(c_1,\dots,c_d,g_s)$ is an element of~\eqref{Equ:ParaRecSol}. Thus
we get $B\neq\{\}$ in step~\eqref{Alg:SolvePRS} and we do not quit
in step~\eqref{Alg:Quit1}. Since $B$ is a $\KK$-basis
of~\eqref{Equ:ParaRecSol}, there is a
$(\kappa_1,\dots,\kappa_n)\in\KK^n$ with
$(c_1,\dots,c_d)=(\kappa_1,\dots,\kappa_n)\,\vect{C}$. By
Lemma~\ref{Lemma:IndefiniteConstraint} we conclude that
$(\kappa_1,\dots,\kappa_n)\vect{\phi}^t=\sigma(g_{s+1})-g_{s+1}$.
Thus the variant $\text{RPT}_r$ is solvable, and the algorithm
cannot return ``No solution'' in step~\eqref{Alg:Quit2}.
Consequently, the output ``No solution'' is not possible, a
contradiction. \qed
\end{proof}

We conclude this section by a concrete example that demonstrates the full flexibility of our refined holonomic machinery to hunt for linear recurrences.

\begin{mexample}\label{Exp:SimpleX}
Given $S(n)=\sum_{k=0}^n \binom{n}{k} X(k)$ with $X(k)=\sum_{j=0}^k \binom{k}{j} S_1({j})^2$, we aim at computing a linear recurrence of the form~\eqref{Equ:Rec}. We start with the \pisiE-field $\dfield{\GG}{\sigma}$ with constant field $\KK=\QQ(n)$ and $\GG=\KK(t)(b)$ where $\sigma(t)=t+1$ and $\sigma(b)=\frac{n-t}{t+1}\,b$.\\
(A) In a first round, we will exploit the recurrence~\eqref{Equ:XOrder4} to set up our h.o.l.\ extension defined $\dfield{\GG}{\sigma}$ (here we can set $\AR=\GG$) and search for a solution $g=g_0\,x_0+g_1\,x_1+g_2\,x_2+g_3\,x_3+g_4$ of Problem~RPT with $g_0,g_1,g_2,g_3\in\GG$ and $g_4$ in $\GG$ or in a properly chosen \rpisiE-extension of $\dfield{\GG}{\sigma}$. First, we will activate Algorithm~\ref{AlgParaTele} with the telescoping strategy $\text{RPT}_1$ for $d=0,1,2,3,\dots$ until we find a recurrence.
Following our algorithm we search for $g_3\in\GG$ and $c_1,\dots,c_d\in\KK$ by solving the following parameterized difference equation
\begin{multline*}
-\tfrac{8 (4+t) (6+t)}{(7+t)^2}\sigma^4(g_3)+\tfrac{4 \big(
        99+45 t+5 t^2\big)}{(6+t)^2}\sigma^3(g_3)-\tfrac{2 (11+3 t) (13+3 t)}{(5+t)^2}\sigma^2(g_3)\\
        +\tfrac{86+49 t+7 t^2}{(4+t)^2}\sigma(g_3)-g_3=c_1\,\tilde{f_1}+\dots+c_d\,\tilde{f_d}
\end{multline*}
where the first six $\tilde{f}_i$ are given by
\begin{align*}
\tilde{f}_1&=-\tfrac{b (-n
+t
) (1
-n
+t
) (2
-n
+t
)}{(1+t) (2+t) (3+t)},&
\tilde{f}_2&=\tfrac{b (1+n) (n
-t
) (-1
+n
-t
)}{(1+t) (2+t) (3+t)},\\
\tilde{f}_3&=\tfrac{b (1+n) (2+n) (n
-t
)}{(1+t) (2+t) (3+t)},
&\tilde{f}_4&=\tfrac{b (1+n) (2+n) (3+n)}{(1+t) (2+t) (3+t)},\\
\tilde{f}_5&=\tfrac{b (1+n) (2+n) (3+n) (4+n)}{(1+t) (2+t) (3+t) (1
+n
-t
)},&
\tilde{f}_6&=\tfrac{b (1+n) (2+n) (3+n) (4+n) (5+n)}{(1+t) (2+t) (3+t) (1
+n
-t
) (2
+n
-t
)}.
\end{align*}
We obtain the first non-trivial solution with $d=5$: the basis $B_5$ of the $\KK$-vector space~\eqref{Equ:ParaRecSol} has dimension $1$ and is given by
$B=\{(c_{1,1},c_{1,2},c_{1,3},c_{1,4},c_{1,5},\gamma_1)\}$
with $c_{1,1}=27 (5+2 n)$,
$c_{1,2}=-\frac{27 (
        28+27 n+6 n^2)}{2 (1+n)}$,
$c_{1,3}=\frac{3 (
        418+544 n+225 n^2+30 n^3)}{2 (1+n) (2+n)}$,
$c_{1,4}=-\frac{414+504 n+187 n^2+22 n^3}{2 (1+n) (2+n)}$,
$c_{1,5}=\frac{(4+n)^2 (3+2 n)}{2 (1+n) (2+n)}$,
and
$\gamma_1=-\frac{b (3+t)^2 (4
+5 t
+n (2+2 t)
)}{2 (1+t) (2+t) (1
+n
-t
)}$. So our hope is that $(c_1,c_2,c_3,c_4,c_5,g_3)$ equals the element of $B_5$. Next, we check if we can determine $g_4\in\GG$ with
\begin{equation}\label{Equ:g4ProbOrder5}
\sigma(g_4)-g_4=\tfrac{b(
        9
        +4 n
        +5 t
        +2 n t
)}{2 (1+t) (2+t) (3+t)}=:\phi.
\end{equation}
Since there is no such solution, we restart our algorithm for $d=6$. This time the $\KK$-vector space~\eqref{Equ:ParaRecSol} has the dimension $2$, i.e., we obtain a basis $B_6$ with two elements (which we do not print here). So we have more flexibility to set up $g_4$. In order to determine $g_4\in\AR$, it must be a solution of $\sigma(g_4)-g_4=\kappa_1\,\phi_1+\kappa_2\,\phi_2$ with
\begin{align*}
 \phi_1&=\tfrac{2 b(
        63
        +46 n
        +8 n^2
        +35 t
        +24 n t
        +4 n^2 t)}{(5+2 n) (1+t) (2+t) (3+t)},
&\phi_2&=-\tfrac{b(
        13
        +4 n
        +7 t
        +2 n t)}{(1+t) (2+t) (3+t) (-1
-n
+t
)};
\end{align*}
for their calculation see~\eqref{Equ:PhiCalculate}.
We find $\kappa_1=\frac{3+2 n}{(5+2 n) (7+2 n)}$, $\kappa_2=-\frac{(3+n) (1+2 n)}{(5+2 n)^2}$,
$g_{4}=-\frac{b (-2 n
-3 t
-2 n t
)}{(5+2 n) (1+t) (2+t) (1
+n
-t
)}$. Combining this solution with entries of $B$ delivers
\begin{multline}\label{Equ:DoubleSumRec5}
108 (1+n) (2+n) (3+2 n) S(n)
-54 (2+n) \big(
        21+30 n+8 n^2\big) S(1+n)\\
+3 \big(
        831+1634 n+795 n^2+114 n^3\big) S(2+n)\\
+\big(
        -1227-2556 n-1095 n^2-134 n^3\big) S(3+n)\\
+\big(
        283+632 n+243 n^2+26 n^3\big) S(4+n)
-(5+n)^2 (1+2 n) S(5+n)
=0.
\end{multline}
\normalsize
Algorithm~\ref{AlgParaTele} with the tactics $\text{RPT}_2$, $\text{RPT}_3$ will deliver the same recurrence. Applying $\text{RPT}_4$ we will obtain for $d=5$ the basis $B_5$ from above and have to find a solution for~\eqref{Equ:g4ProbOrder5}. Since there is no solution $g_4\in\GG$ (and the tactics from $\text{RPT}_2$, $\text{RPT}_3$ fail), we continue and construct the \sigmaE-extension $\dfield{\GG[\tau]}{\sigma}$ of $\dfield{\GG}{\sigma}$ with $\sigma(\tau)=\tau+\phi$ and get the solution $g_4=\tau$. Gluing all the building blocks together, \texttt{Sigma} delivers
\begin{multline}\label{Equ:RecWithDefiniteSum2}
54 (1+n) (2+n) (5+2 n) S(n)
-27 (2+n) \big(
        28+27 n+6 n^2\big) S(1+n)\\
+3 \big(
        418+544 n+225 n^2+30 n^3\big) S(2+n)
-\big(
        414+504 n+187 n^2+22 n^3\big) S(3+n)\\
+(4+n)^2 (3+2 n) S(4+n)
=2 n
+(1+n) (2+n)
\sum_{i=0}^n \tfrac{\binom{n}{i} (4
+5 i
+2 n
+2 i n
)}{(1+i) (2+i) (1
-i
+n
)}.
\end{multline}
We remark that the found sum on the right hand side can be turned to an expression in terms of indefinite nested objects (like for the sum~\eqref{Equ:RHSSum}) and the right hand side can be simplified to $(1+2 n) 2^{2+n}$.\\
(B) In the second round, we will exploit the recurrence~\eqref{Equ:XOrder2} and set up a h.o.l.\ extension defined over a properly chosen \rpisiE-extension $\dfield{\AR}{\sigma}$ of $\dfield{\GG}{\sigma}$. In this setting we search for a solution $g=g_0\,x_0+g_1\,x_1+g_2$ of Problem~RPT with $g_0,g_1\in\GG$ and $g_2$ in $\AR$ or in a properly chosen \rpisiE-extension of $\dfield{\AR}{\sigma}$. If we apply tactic~$\text{RPT}_1$ or $\text{RPT}_2$, we will get~\eqref{Equ:DoubleSumRec5}. However, if we apply $\text{RPT}_3$, we find
\begin{multline}\label{Equ:RhoSumOutput}
-36 (1+n)^2 (2+n) (3+n) S(n)
+6 (1+n) (2+n) (3+n) (12+7 n) S(1+n)\\
+2 (-19-8 n) (1+n) (2+n) (3+n) S(2+n)
+2 (1+n) (2+n) (3+n)^2 S(3+n)\\
=-2^{3+n} (1+n)^2
+2\ 3^{2+n} (1+n) (3+2 n).
\end{multline}
Finally, if we activate tactic $\text{RPT}_4$ in \texttt{Sigma}, we end up at the recurrence
\begin{multline*}
-9 (1+n) S(n)
+3 (3+2 n) S(1+n)
+(-2-n) F(2+n)\\
=\tfrac{2^{1+n} (4+3 n)}{(1+n) (2+n)}
-\tfrac{3^{1+n} (5+4 n)}{(1+n) (2+n)}
+2^{1+n} S_1({n})
-2^{1+n} S_1\big({{\tfrac{3}{2}},n}\big).
\end{multline*}
Note that we obtained in both cases first a recurrence where on the right hand side definite sums pop up which afterwards are simplified to indefinite versions.\\
(C) In a third round, one can use the zero-order recurrence~\eqref{Equ:XOrder0} following the standard \texttt{Sigma}-approach~\cite{Schneider:13} and can apply purely the tools from Subsection~\ref{Subsec:PureRPiSi} (see Remark~\ref{Remark:RPTVariants}). In contrast to the variants (A) and (B), these calculations are more involved since they have to be carried out within a much larger \rpisiE-extension.\\
Solving any of the found recurrences in terms of d'Alembertian solutions yields
\begin{multline*}
S(n)=3^n\big(
        -2 S_1({n}) S_1\big({{\tfrac{2}{3}},n}\big)
        -2 S_2\big({{\tfrac{2}{3}},n}\big)
        -S_{1,1}\big({{\tfrac{2}{3},\tfrac{3}{2}},n}\big)\\
        +3 S_{1,1}\big({{\tfrac{2}{3},1},n}\big)
        +S_1({n})^2
        +S_2({n})
\big).
\end{multline*}
\end{mexample}

Summary: we provided different holonomic summation tactics in the context of \rpisiE-extensions to find linear recurrences.
The smaller the obtained recurrence order is, the more the underlying difference ring algorithms are challenged to handle many \rpisiE-extensions. Conversely, the higher the recurrence order is, the larger will be the computed coefficients of the recurrence and thus the underlying arithmetic operations get more involved.

\section{A multi-sum method to determine recurrences}\label{Sec:MultiSumApproach}


We aim at computing a recurrence of an $m$-fold definite nested multi-sum
\begin{equation}\label{Equ:HSum}
S(n) = \sum_{k_1=\alpha_1}^{L_1(n)}h_1(n,k_1)\sum_{k_2=\alpha_2}^{L_2(n,k_1)}h_2(n,k_1,k_2) \cdots \sum_{k_m=\alpha_m}^{L_m(n,k_1,\ldots, k_{m-1})} h_m(n,k_1 ,\ldots , k_m)
\end{equation}
where for $1\leq i\leq m$ the following holds: $\alpha_i\in\NN$, $L_i(n,k_1,\dots,k_{i-1})$ stands for an integer linear expression or equals $\infty$, and $h_i(n,k_1,\dots,k_i)$
is an expression in terms of indefinite nested sums over hypergeometric products w.r.t.\ the variable $k_i$.

\begin{mdefinition}\label{Def:IndefiniteNested}
Let $f(k)$ be an expression that evaluates at non-negative
integers (from a certain point on) to elements of a field $\KK$. $f(k)$
is called an \emph{expression in terms of indefinite nested sums over hypergeometric products  w.r.t.\ $k$}  if it is composed of elements from the rational function field $\KK(k)$, by the three operations
($+,-,\cdot$), by \emph{hypergeometric products} of the form $\prod_{j=l}^k h(j)$ with $l\in\NN$ and a rational function $h(t)\in\KK(t)\setminus\{0\}$, and by sums of the form $\sum_{j=l}^k F(j)$ with $l\in\NN$ and where
$F(j)$, being free of $k$, is an expression in terms of indefinite nested sums over hypergeometric products w.r.t.\ $j$.
\end{mdefinition}

\noindent
For this task we will improve substantially the multi-sum approach introduced in~\cite{Schneider:05a} by exploiting our new difference ring machinery from Section~\ref{DFTheory}. More precisely, we will process the sums in~\eqref{Equ:HSum} from inside to outside and will try to compute for each sub-sum $X(\vect{n},k)$ a refined holonomic system\footnote{Also in ~\cite{CHYZAK2000115} coupled systems are constructed to handle multi-sums. Here we restrict to a special form so that the full power of our tools from Section~\ref{Sec:DRTools} can be applied without using any Gr\"ober bases or uncoupling computations. In particular, the recurrences can have inhomogeneous parts which can be represented in \pisiE-fields and \rpisiE-extensions. Also the coefficients could be represented in general \pisiE-fields (see~\cite{Schneider:05a}), but we will skip this more exotic case.} w.r.t.\ $k$.

\begin{mdefinition}\label{Def:RefinedSys}
Consider a multivariate sequence $X(\vect{n},k)$ with the distinguished index $k$ and further indices  $\vect{n}=(n_1,\dots,n_u)$ and let $\vect{e}_i$ be the $i$th unit vector of length $u$. A refined holonomic system for $X(\vect{n},k)$ w.r.t.\ $k$ is a set of equations of the form
\begin{align}\label{Equ:PureRec}
X(\vect{n},k+s+1)&=A_0(\vect{n},k)\,X(\vect{n},k)+\dots+A_{s}(\vect{n},k)\,X(\vect{n},k+s)+A_{s+1}(\vect{n},k),\\
\label{Equ:HookRec}
X(\vect{n}+\vect{e}_i,k)&=A^{(i)}_0(\vect{n},k)\,X(\vect{n},k)+\dots+A^{(i)}_{s}(\vect{n},k)\,X(\vect{n},k+s)+A^{(i)}_{s+1}(\vect{n},k)
\end{align}
with $1\leq i\leq u$
which holds within a certain range of $k$ and $\vect{n}$ and
where the $A_j(\vect{n},k)$ and $A_j^{(i)}(\vect{n},k)$ with $0\leq j\leq s$ and $1\leq i\leq u$ are rational functions in $K(\vect{n},k)$ for some field $K$ and the $A_{s+1}(\vect{n},k)$ and $A_{s+1}^{(i)}(\vect{n},k)$ for $1\leq i\leq u$ are indefinite nested sums over hypergeometric products w.r.t.\ $k$.
\end{mdefinition}


\noindent \textbf{Base case.} We process the trivial sum $X(n,k_1,\dots,k_{m})=1$ and can construct the refined holonomic system $X(n+1,k_1,\dots,k_{m})=X(n,k_1,\dots,k_{m})$ and $X(n,k_1,\dots,k_i+1,\dots,k_{m})=X(n,k_1,\dots,k_i,\dots,k_{m})$ for all $1\leq i\leq m$.

\medskip


\noindent Now suppose that we succeeded in treating the sum
\begin{equation}\label{Equ:LastSummandRewritten}
X(n,k_1,\dots,k_{u})=\sum_{k_{u-1}=\alpha_{u-1}}^{L_{u+1}(n,k_1,\dots,k_{u})}h_{u-1}(\dots) \cdots \sum_{k_m=\alpha_m}^{L_m(n,k_1,\ldots, k_{m-1})} h_m(n,k_1 ,\ldots , k_m).
\end{equation}
For convenience, set $\vect{n}=(n_1,\dots,n_u):=(n,k_1,\dots,k_{u-1})$ and $k=k_{u}$; further set $\tilde{\vect{n}}=(n_1,\dots,n_{u-1})$.
By assumption we computed a refined holonomic system for $X(\vect{n},k)=X(\tilde{\vect{n}},n_u,k)$ w.r.t.\ $k$  as given in Definition~\ref{Def:RefinedSys}.\\
If $u=0$, we are done. Otherwise we proceed as follows.

\medskip

\noindent \textbf{Recursion step.}
Consider the next sum
$$\tilde{X}(\vect{n})=\tilde{X}(\tilde{\vect{n}},n_u)=\sum_{k=\alpha_u}^{L_u(\vect{n})}F(\vect{n},k)$$
with  $F(\vect{n},k)=h_u(\vect{n},k)X(\vect{n},k)$. Then we aim at computing a refined holonomic system for $\tilde{X}(\vect{n})$  w.r.t.\ $n_u$.
Namely, set $F_i(k)=F(\vect{n}+(i-1)\,\vect{e}_u,k)=F(\tilde{\vect{n}},n_u+i-1,k)$. Then using the rewrite rules~\eqref{Equ:PureRec} and ~\eqref{Equ:HookRec} we can write $F_i(k)$ as
$$F_i(k)=F(\vect{n}+(i-1)\,\vect{e}_u,k)=F_{i,0}(k)\,X(k)+\dots+F_{i,s}(k)\,X(k+s)+F_{i,s+1}(k)$$
where for $1\leq i\leq d$ and $0\leq j\leq s+1$  the $F_{i,j}(k)$ are indefinite nested sums over hypergeometric products w.r.t.\ $k$. Given this form, we try to construct an \rpisiE-extension $\dfield{\AR}{\sigma}$ of a \pisiE-field $\dfield{\GG}{\sigma}$ with the following properties:
we can rephrase the $A_j(k)$ from~\eqref{Equ:PureRec}
with $1\leq j\leq s$ by $a_{j}$ in $\GG$ and $A_{s+1}$ by $a_{s+1}$ in $\AR$, and simultaneously, we can rephrase the $F_{i,j}(k)$ with $1\leq i\leq d$ and $0\leq j\leq s$ by $f_{i,j}$ in $\GG$ and the $F_{i,s+1}$ with $1\leq i\leq d$ by $f_{i,s+1}$ in $\AR$.

\begin{mremark}\label{Remark:SpecialCaseForSummand}
(1) Consider the special case $u=m$. Looking at the base case, we get $s=0$ with $a_0=1$ and $a_1=0$. Further,
$F_1(\vect{n},k)=h_m(\vect{n},k)$ is given in terms of indefinite nested sums over hypergeometric products, and also the shifted versions $F_i(\vect{n},k)=h_m(\vect{n}+(i-1)\vect{e}_u,k)$ for $i\geq2$ are again from this class. All these objects can be rephrased in one common \rpisiE-extension $\dfield{\AR}{\sigma}$ of the rational difference field $\dfield{\GG}{\sigma}$ with $\GG=\KK(t)$ and $\sigma(t)=t+1$ using the algorithms from~\cite{Schneider:05c,DR1,DR2,OS:17}. In a nutshell: for this special case the desired construction is always possible.\\
(2) If $h_u(\vect{n},k)\in K(\vect{n},k)$ (for some field $K$), then $F_{i,j}(k)\in K(\vect{n},k)$ for all $1\leq i\leq d$ and $0\leq j\leq s$. In addition, $A_j(\vect{n},k)\in K(\vect{n},k)$ for all $0\leq j\leq s$ by our recursive construction. Further, $F_{i,s+1}(\vect{n},k)$ with $1\leq i\leq d$ and $A_{s+1}(\vect{n},k)$ are indefinite nested sums over hypergeometric products w.r.t.\ $k$.
Hence by using our tools from~\cite{Schneider:05c,DR1,DR2,OS:17}, we can accomplish this construction.\\
If $h_u(\vect{n},k)$ ($1\leq u<m$) is more involved, we refer to part (3) of Remark~\ref{Remark:TechnicalDetails} below.
\end{mremark}

\noindent If this rephrasing in $\GG$ and $\AR$ is possible, take the h.o.l.\ $\dfield{\HH}{\sigma}$ of $\dfield{\AR}{\sigma}$ with $\HH=\AR[x_0,\dots,x_s]$ with~\eqref{HOEShift}. In other words, we model $F_i(k)$ by~\eqref{Equ:fiInX}.
Now we activate our Algorithm~\ref{AlgParaTele} by choosing an appropriate tactic $\text{RPT}_r$ with $r\in\{1,2,3,4\}$: for $d=0,1,2,\dots$ we check with the input $\vect{f}=(f_1,\dots,f_{d})$
if we find a solution for Problem~$\text{RPT}_r$. If we succeed for $d$ ($d$ is minimal for a given tactic) and rephrase the found solution in terms of indefinite nested sums and products, we obtain the summand recurrence~\eqref{eq:ptcert} and summing this equation over the summation range\footnote{If there are exceptional points within the summation range, we refer to Subsection~\ref{Sec:ExceptionalPoints}. Further, if the upper bound is $\infty$, limit computations are necessary. For wide classes of indefinite nested sums asymptotic expansions can be computed~\cite{ABS:11,Ablinger:12,Ablinger:2013cf,ABRS:14} that can be used for this task.} yields a recurrence of the form~\eqref{Equ:PureRec} for the next sum $\tilde{X}(\tilde{\vect{n}},n_u)$.\\
Similarly, choose $i$ with $1\leq i<u$. Then we can set
$F^{(i)}_0=F(\vect{n}+\vect{e}_i,k)$ and using the rewrite rules ~\eqref{Equ:PureRec} and ~\eqref{Equ:HookRec} we obtain
$$F_0^{(i)}(k)=F(\vect{n}+\vect{e}_i,k)=F^{(i)}_{0}(k)\,X(k)+\dots+F^{(i)}_s(k)\,X(k+s)+F^{(i)}_{s+1}(k)$$
where the $F^{(i)}_{j}(k)$ are indefinite nested sums over hypergeometric products w.r.t.\ $k$. As above, we try to represent these elements by $f_0^{(i)}=f^{(i)}_0\,x_0+\dots+f^{(i)}_s\,x_s+f^{(i)}_{s+1}$ with $f^{(i)}_j\in\GG$ for $0\leq j\leq s$ and $f^{(i)}_{s+1}\in\AR$ in an \rpisiE-extension $\dfield{\AR}{\sigma}$ of a \pisiE-field $\dfield{\GG}{\sigma}$.
Now we activate again Algorithm~\ref{AlgParaTele} with the input $\vect{f}=(f_0^{(i)},f_1,\dots,f_{\delta})$ for $\delta=0,1,\dots$. In all our applications we have been successful for a $\delta$ with $\delta<d$ by
choosing one of the tactics $\text{RPT}_j$ with $j\in\{1,2,3,4\}$ (usually, with the tactic $\text{RPT}_r$ that lead to~\eqref{Equ:PureRec}). In other words, reinterpreting this solution in terms of indefinite nested sums and products and summing the found equation over the summation range will produce a recurrence of the form~\eqref{Equ:HookRec} for $\tilde{X}(\tilde{\vect{n}},n_u)$.

Performing this calculation for all $i$ with $1\leq i<u$ yields a system of recurrences for $\tilde{X}(\tilde{\vect{n}},n_u)$ of the form~\eqref{Equ:PureRec} and~\eqref{Equ:HookRec}. To turn this to a refined holonomic system, one has to face an extra challenge. The inhomogeneous sides $A_{s+1}(k)$ and $A_{s+1}^{(i)}(k)$ often contain definite sums (see, e.g., the recurrences~\eqref{Equ:XOrder2Raw} and~\eqref{Equ:RecWithDefiniteSum2}). To rewrite them to indefinite nested versions (which are expressible in an \rpisiE-extension) further symbolic simplifications are necessary; see Subsection~\ref{Subsec:DefiniteSums} below.
If this is not possible, our method fails. Otherwise, this completes the recursion step of our method.

\begin{mremark}\label{Remark:TechnicalDetails}
(1) If a refined holonomic system with $s=0$ arises in one of these recursion steps (this is in particular the case if we treat the first summation), Algorithm~\ref{AlgParaTele} boils down to solve Problem~RPT in $\dfield{\AR}{\sigma}$; compare also Remark~\ref{Remark:ZeroOrder}.\\
(2) If the expression $\tilde{X}(\vect{n})$ in the recursion step is free of $n_i$ ($1\leq i\leq u$), one gets trivially $\tilde{X}(\vect{n}+\vect{e}_i)=\tilde{X}(\vect{n})$.\\
(3) Given~\eqref{Equ:HSum}, the summands $h_i(n,k_1,\dots,k_i)$ with $i<m$ (i.e., not the innermost summand $h_m$) might introduce complications. The indefinite nested sums over hypergeometric products w.r.t.\ $k_i$ in the $h_i$  and their shifted versions in the parameters $(n,k_1,\dots.k_{i-1})$ must be encoded in a \pisiE-field $\dfield{\GG}{\sigma}$; see Remark~\ref{Remark:SpecialCaseForSummand}. If this is not possible, our method fails. If it works, Problem~PRS has to be solved in $\dfield{\GG}{\sigma}$. Hence the \pisiE-field should be composed only by a reasonable sized set of generators to keep the algorithmic machinery efficient.
Conversely, if one sets $h_1=\dots=h_{m-1}=1$ in~\eqref{Equ:HSum} and moves all summation objects into $h_m$, one can choose for $\dfield{\GG}{\sigma}$ the rational difference field; see Remark~\ref{Remark:SpecialCaseForSummand}. However, in this case the inhomogeneous parts of the refined holonomic system will blow up. In our experiments we found out that choosing $h_i$ as a hypergeometric product (that can be formulated in a \pisiE-field) was a reasonable trade-off to gain speed up and to keep the \pisiE-field simple; see Example~\ref{Exp:SimpleX} for a typical application.
\end{mremark}

Our machinery works also for sums~\eqref{Equ:HSum} where the $h_i$ depend on mixed multi-basic hypergeometric products. This means that in Definition~\ref{Def:IndefiniteNested} one also allows products of the form $\prod_{j=l}^kf(j,q_1^j,\dots,q_e^j)$ where $f(t,t_1,\dots,t_e)$ is a rational function.
The only extra adaption is to take as ground field instance (3) of Example~\ref{Exp:PiSi}.





\subsection{Illustrative examples}\label{Exp:TripleSum}

Our working example (see Examples~\ref{Exp:PiSi} and~\ref{Exp:SimpleX}) follows precisely the above multi-sum method.
Namely consider our sum $S(n)=\sum_{k=0}^n \binom{n}{k}X(n,k)$ with $X(n,k)=\sum_{j=0}^k \binom{k}{j} S_1({j})^2$.
We worked from inside to outside and computed a refined holonomic system for each summand. First, we took the inner sum $X(k)=X(n,k)$ and computed a recurrence purely in $k$ demonstrating our different telescoping strategies. Since $X(k)$ is free of $n$, we get trivially the recurrence
$X(n+1,k)=X(n,k).$
Afterwards, we applied our multi-sum machinery to the second sum: namely, as worked out in Example~\ref{Exp:SimpleX} we computed a recurrence of $S(n)$ by exemplifying our different summation tactics.

Now let us turn to a~\texttt{Mathematica}--implementation of the refined holonomic approach called~\texttt{RhoSum}.  It is built on top of~\texttt{Sigma},~\texttt{HarmonicSums}~\cite{Ablinger:12} and~\texttt{EvaluateMultiSums}~\cite{Schneider:13}. The first step is to load these packages,
\begin{mma}
\In << Sigma.m \vspace*{-0.06cm}\\
\Print Sigma - A summation package by Carsten Schneider
\copyright\ RISC\\
\In << HarmonicSums.m \vspace*{-0.06cm}\\
\Print HarmonicSums by Jakob Ablinger -- \copyright\ RISC\\
\In << EvaluateMultiSums.m \vspace*{-0.06cm}\\
\Print EvaluateMultiSums by Carsten Schneider -- \copyright\ RISC\\
\In << RhoSum.m \vspace*{-0.06cm}\\
\Print RhoSum by Mark Round -- \copyright\ RISC\\
\end{mma}
\noindent By loading these packages one obtains recurrence finding and solving tools from \texttt{Sigma}, special function algorithms for indefinite nested sums~\cite{ABS:11,Ablinger:12,Ablinger:2013cf,ABRS:14} from \texttt{HarmonicSums}, and summation technologies from \texttt{EvaluateMultiSums} and finally the refined summation package itself, \texttt{RhoSum}.  Then with a single command the above method is applied to our double sum to deliver a recurrence.
\begin{mma}
\In FindRecurrence[\binom{n}{k}\binom{k}{j} S_1[{j}]^2,\{\{j,0,k\},\{k,0,n\}\},\{n\},\{0\},\{\infty\}]\\
\Out -36 (1+n)^2 (2+n) (3+n) nSUM[n]
+6 (1+n) (2+n) (3+n) (12+7 n) nSUM[1+n]\newline
\hspace*{0.3cm}+2 (-19-8 n) (1+n) (2+n) (3+n) nSUM[2+n]\newline
\hspace*{0.7cm}+2 (1+n) (2+n) (3+n)^2 nSUM[3+n]==
-2^{3+n} (1+n)^2
+2\ 3^{2+n} (1+n) (3+2 n)
\\
\end{mma}
\noindent Internally, \texttt{RhoSum} used up to a certain complexity the subroutines of \texttt{Sigma} with the tactic $\text{RPT}_3$ (see Theorem~\ref{Thm:OptimalDF}) and delivers the recurrence~\eqref{Equ:RhoSumOutput}. If one wants to solve the recurrence in addition in terms of d'Alembertian solutions (in case this is possible), one can execute the command
\begin{mma}
\In FindSum[\binom{n}{k}\binom{k}{j} S_1({j})^2,\{\{j,0,k\},\{k,0,n\}\},\{n\},\{0\},\{\infty\}]\\
\Out 3^n\big(
        -2 S_1[{n}] S_1\big[{{\tfrac{2}{3}},n}\big]
        -2 S_2\big[{{\tfrac{2}{3}},n}\big]
        -S_{1,1}\big[{{\tfrac{2}{3},\tfrac{3}{2}},n}\big]
        +3 S_{1,1}\big[{{\tfrac{2}{3},1},n}\big]
        +S_1[{n}]^2
        +S_2[{n}]\big)\\
\end{mma}

We will concentrate on the slightly more involved triple sum
\begin{equation*}
S(N)= \sum_{n=0}^N \overbrace{\sum_{k=0}^{n+N}
\underbrace{\sum_{j=0}^k S_1(j)\binom{n+N}{j}^2}_{=:c_{N,n,k}}}^{=:b_{N,n}}
\end{equation*}
in order to outline all steps of our multi-sum method.
Our aim is to compute a refined holonomic system for the complete multi-sum $S(N)$.  This refined holonomic system is particularly simple, it consists of just one recurrence of shifts in $N$. To compute the recurrence our algorithm is to encode the summand $b_{N,n}$ in to a refined holonomic system too. This is a system of two recurrences involving shifts in $N$ and $n$.  Again this will be done by encoding the summand $c_{N,n,k}$
in yet another refined holonomic system.  This is the base case because we can compute with the summand explicitly.  Using~\texttt{Sigma} the refined holonomic system
\begin{align*}
-c_{N,n,k} + c_{N,n,k+1}
=\binom{n
+N
}{k}^2 \big(
        \tfrac{(k
        -n
        -N
        )^2}{(1+k)^3}&
        +\tfrac{(k
        -n
        -N
        )^2}{(1+k)^2}S_1({k})
\big)\\
2 (1
+2 n
+2 N
) c_{N,n,k}
-(1
+n
+N
) c_{N,n+1,k}
&=E(N,n,k)\\
2 (1
+2 n
+2 N
) c_{N,n,k}
-(1
+n
+N
) c_{N+1,n,k}
&=E(N,n,k)
\end{align*}
can be computed with
$$E(N,n,k)=
\big(
        \tfrac{1
        -3 k
        +4 n
        +4 N
        }{1
        +n
        +N
        }
        +(1
        -2 k
        +3 n
        +3 N
        ) S_1({k})
\big) \tbinom{n
+N
}{k}^2
+\tfrac{(-1
-4 n
-4 N
)}{1
+n
+N
}(1+ s(k))$$
which contains the extra sum
$ s(k) = \sum_{i=1}^k \tbinom{n
+N
}{i}^2.$
Next we use this system to compute a refined holonomic system for the sequence $b_{N,n}$.  Using~\texttt{Sigma} we get a recurrence purely shifted in $n$:
\begin{multline*}
2 (-1
+n
+N
) (1
+2 n
+2 N
) b_{N,n}
-(-2
+n
+N
) (1
+n
+N
) b_{N,n+1}\\
=2 (1
+2 n
+2 N
) c_{N,n,0}
+4 (1
+2 n
+2 N
) c_{N,n,N+n}\\
+\tfrac{(
        5
        +20 n
        -7 n^2
        -4 n^3
        +20 N
        -14 n N
        -12 n^2 N
        -7 N^2
        -12 n N^2
        -4 N^3
)}{2 (1
+n
+N
)}(1+s(n
+N
))
.
\end{multline*}
Notice that the middle line contains ``telescoping points'' (see also Subsection~\ref{Subsec:DefiniteSums}) and $s(n+N)$ turns to a definite sum (the integer parameters $N$ and $n$ arise inside the sum and at the upper bound). These give three new summation problems. $c_{N,n,0}=0$ is trivial, while the remaining sums can be treated similar to the sum in Example~\ref{Exp:PiSi}. Namely, we get
\begin{align*}
c_{N,n,N+n} &= \sum_{j=0}^{n+N} S_1(j)\tbinom{n+N}{j}^2  = \frac{3}{2} \frac{(2 n
+2 N
)! S_1({n+N})}{((n
+N
)!)^2}
-\frac{(2 n
+2 N
)!}{((n
+N
)!)^2} \sum_{j=1}^{n+N} \frac{1}{2j-1},\\
 s(n+N) &= -1
+\frac{(2 n
+2 N
)!}{((n
+N
)!)^2}.
\end{align*}
The result contains a cyclotomic harmonic sum~\cite{ABS:11}. Replacing these definite sums with their indefinite nested sum representations leads to the final recurrence for $b_{N,n}$ purely with shifts in $n$. There is also a recurrence shifted in $N$ (note that $n$ and $N$ are symmetric) and we end up at the refined holonomic system
\begin{align*}
2 (-1
+n
+N
) (1
+2 n
+2 N
) b_{N,n}
-(-2
+n
+N
) (1
+n
+N
) b_{N,n+1}&=r(N,n),\\
2(-1
+n
+N
) (1
+2 n
+2 N
) b_{N,n}
-(-2
+n
+N
) (1
+n
+N
) b_{N+1,n}&=r(N,n)
\end{align*}
with the same right hand side
\begin{align*}
r(N,n)=&4 (1
+2 n
+2 N
) \big(
        \frac{3 (2 n
        +2 N
        )! S_1({n+N})}{2 ((n
        +N
        )!)^2}
        -\frac{(2 n
        +2 N
        )! \sum_{j=1}^{n+N} \tfrac{1}{2j-1}}{((n
        +N
        )!)^2}
\big)\\
&+\tfrac{\big(
        5
        +20 n
        -7 n^2
        -4 n^3
        +20 N
        -14 n N
        -12 n^2 N
        -7 N^2
        -12 n N^2
        -4 N^3
\big)}{2 (1
+n
+N
)}\frac{(2 n
+2 N
)!}{(n
+N
)!^2}.
\end{align*}
Finally, this system can be used to compute a recurrence for the entire multi-sum.  Using~\texttt{Sigma} one obtains
\begin{align*}
&\tfrac{3 (1+4 N) (7+9 N)}{(-1+N) (1+N) (1+2 N)}S_1({2 N})
+\tfrac{(
        43+378 N+527 N^2-312 N^3-828 N^4-288 N^5)}{4 (-1+N) (1+N)^2 (1+2 N)^2}\frac{(4 N)!}{(2 N)!^2}+b_{N,0}\\
&
+\tfrac{(
        1-3 N-38 N^2-40 N^3) b_{N,N}}{(-1+N) (1+N) (1+2 N)}
-\tfrac{2 (1+4 N) (7+9 N)}{(-1+N) (1+N) (1+2 N)}\frac{(4 N)!}{(2 N)!^2} \sum_{j=1}^{2N}\tfrac{1}{2j-1}
 = S(N)-S(N+1).
\end{align*}
There are two telescoping points $b_{N,N}$ and
$b_{N,0}$ to evaluate which turn out to be two double sums. Applying our machinery again to these sums gives recurrences in $N$ and solving them produces closed form solutions in terms of indefinite nested sums. Plugging these simplifications into the telescoping points provides the final result: a recurrence for $S(N)$ in terms of indefinite nested sums. We remark that this recurrence can be also solved in terms of indefinite nested sums over hypergeometric products, but the result is too big to be printed here.




\subsection{\label{TechRemarks}Implementation remarks}

This section contains a discussion of various technical components, knowledge of which is required for an efficient implementation of the underlying machinery in \texttt{RhoSum}.
Some remarks refer to how the recurrence should be computed and handled before being returned by the recurrence finding technology.  As such, in terms of our multi-sum approach described in the beginning of Section~\ref{Sec:MultiSumApproach}, these comments fit inside the calls to the recurrence finding technology.

Usually recurrence finding technologies are very costly. In general, a multi-summation algorithm based on refined difference field theory will compute up to $m(m+1)/2$ recurrences in an $m$-fold sum.  This makes controlling the individual recurrences very important because if  any individual recurrence is too large in size or the underlying difference ring consists of too many \rpisiE-monomials then the inherently high cost of recurrence finding technologies may easily lead to the entire multi-sum problem becoming intractable.  Notice also that because recurrences are computed from sums of lower depth, efficiency issues can become cumulative; a recurrence which is not expressed in a simple form is likely to lead to longer computation times when used as an input for another calculation.  This further serves to highlight the importance of the technical details of recurrence computations.  We will discuss some of the specific issues that fit into this central problem.

\subsubsection{Managing recurrence computation time}
When computing a recurrence, one must define the desired type of recurrence.  The definition corresponds to how one constructs $g_{s+1}$ in Theorem~\ref{Thm:OptimalDF}.  A configuration that searches for minimal order recurrences translates to applying tactic $\text{RPT}_4$ where $d$ is minimal. Such an approach is, relatively speaking, cheap to compute; the linear system one must solve for the homogeneous part is of minimal size and one takes the first available solution then extends the ring to get an inhomogeneous solution.  The potential penalty is to adjoin a \sigmaE-extension which might be rather complicated which afterwards has to be converted to an expression in terms of indefinite nested sums. At the other extreme, one can use tactic $\text{RPT}_1$, i.e., to relax the condition of minimal order and try to compute $g_{s+1}$ in the difference ring that one uses to describe the input problem -- and if this fails to increase the recurrence order $d$ of the parameterized telescoping problem.
When working with harder sums the different approaches, including tactics $\text{RPT}_2$ and $\text{RPT}_3$, may have very different computation times. The different possibilities are carried out in Example~\ref{Exp:SimpleX}.\\
There are several reasonable heuristics, one could choose a single methodology for all recurrence computations.  In the case of particle physics sums this can be useful.  In fact, if a minimal order approach is taken for sums originating from particle physics, then the computations are likely to be representative of an optimum balance of the two methods.  This is a strong motivation for pursuing refined holonomic summation.  It offers the best approach to particle physics sums as compared to other techniques.  Always applying a non-minimal order approach is likely to require computations that are not feasible with modern computer power.  One could also choose to switch between the two methods.  For example a crude but simple approach would be to adopt some non-minimal order approach and if the search is yet to return a result after a given time limit one switches to a minimal order approach.  Another option would be to limit the order the non-minimal search takes place over.  When that order has been exceeded one would then switch to a minimal order approach.  Implementing both a time and recurrence order limit is recommended to avoid the scenario that a low but non-minimal order recurrence is very slow to compute and so must be avoided but at the same time allow the implementation to find simple recurrences of relatively high order.



\subsubsection{Definite sums inside of recurrences}\label{Subsec:DefiniteSums}

As already observed in part~(3) of Remark~\ref{Remark:TechnicalDetails}, one has to deal with definite sums that arise within the calculation of recurrence relations. Namely, given a summand recurrence~\eqref{eq:ptcert} for properly chosen summands $F_i(k)$, one obtains a recurrence relation~\eqref{Equ:Rec}
where $h(n)$ comes from $G(L(n)+1) - G(l)$ which either evaluates nicely or otherwise turns to definite sums. More precisely, one either obtains the so-called telescoping points $X(l)$ or $X(L(n)+i)$ with $i\in\NN$ or one gets definite sums coming from certain \sigmaE-extensions.
In both cases, these sums are simpler summation problems: in the first case, they are simpler than the main multi-sum because they are at specific values, in the second case they come from our \rpisiE-extensions which can be formulated in a simpler \rpisiE-extension (the summand can be formulated in a smaller difference ring).  In other words, we have to solve simpler summation problems and the resulting recursive calls of our algorithms (i.e., calling \texttt{FindSum}) will eventually terminate. We remark that higher-order recurrences lead to larger numbers of telescoping points (tactic $\text{RPT}_1$) but more involved \rpisiE-extensions (tactic $\text{RPT}_4$) might also lead to more complicated definite sums. Here tactics $\text{RPT}_2$ and $\text{RPT}_3$ can be an interesting alternative to reduce the calculation time concerning the treatment of extra definite sums and avoiding any slow down of our refined holonomic summation implementation.  With modern computers it is likely however that the telescoping points can be computed simultaneously by using parallelization.

\subsubsection{Exceptional points}\label{Sec:ExceptionalPoints}

It may be that the function $G(k)$ computed in solving the parameterized telescoping problem (compare to~\eqref{eq:ptcert}) is not defined for some values in the range of summation.  Thus when one tries to sum over the expression to obtain~\eqref{eq:ptproblem}, one encounters ill-defined expressions even though the original summation problem is well-defined at that point. Usually such exceptional points restrict the summation range by a difference of one or two, should they occur at all. E.g., consider the sum $b_{n,k}=\sum_{j=0}^k c_{n,k,j}$ for which we want to compute a refined holonomic system, and suppose that we find a refined holonomic system for $c_{n,k,j}$ which is only valid within the range $j=j_1,\dots,k-j_2$ for some $j_1,j_2\in\NN$. If one wishes to continue with a refined approach, there are two options.  One could accept the restricted ranges returned by the summation technology and continue without making any adjustments.  Then the final expression will be valid for a different sum which is contained within the original sum and it is most likely that only significantly simpler sums are required to solve the entire problem.  However by disturbing the structure of the multi-sum, many unwanted, and possibly hard, sums might not cancel and one is faced with extra work to treat these sums.
Alternatively one can compute the values of the exceptional points and compensate in the problem.  To do this consider the rewriting
\begin{equation*}
b_{n,k}= \sum_{j=0}^{k} c_{n,k,j}=
\sum_{j=0}^{j_1-1} c_{n,k,j} +\sum_{j=k-j_2+1}^{k} c_{n,k,j} +\sum_{j=j_1}^{k-j_2} c_{n,k,j}=K+ \sum_{j=j_1}^{k-j_2} c_{n,k,j}.
\end{equation*}
The expression for $K$ is just given by two definite sums, which are often easy to handle.  Within this approach there is a subtlety as to where one places $K$ in the multi-sum expression.  The choices are
\begin{equation*}
a_n= \sum_{k=0}^{n} \Big(K+\sum_{j=j_1}^{k-j_2} c_{n,k,j}\Big)\quad\text{ or }\quad
a_n = \sum_{k=0}^{n} \sum_{j=j_1}^{k-j_2} \left(\tfrac{K}{k-j_2-j_1+1}+c_{n,k,j} \right).
\end{equation*}
In general, our heavy calculations coming from particle physics gave the experience that the second strategy is more preferable: within the summand further cancellations arise and the processing of the summations turn out to be easier.




\section{\label{Examples}Examples from elementary particle physics}

Perturbative calculations in quantum field theory lead to various summation problems~\cite{BKSF:12}, and one of the challenges is to find recurrence relations of a certain order and polynomial degree, where the polynomials
contain an integer variable $N$ and a series of parameters. One of which is the dimensional parameter $\varepsilon =
D - 4$, which is required to handle divergences in the Feynman diagrams.  This introduces a small parameter $\varepsilon >0$ inside of the sum~\eqref{Equ:HSum}.
Some of our results reproduce calculations that have only recently entered the particle physics literature~\cite{Physics1,Physics2,Physics3}.
In some very rare cases one can apply directly our method \texttt{FindRecurrence} to compute a recurrence for $S(n)$ or to apply \texttt{FindSum} to compute a closed form in terms of indefinite nested sums. In such cases the derived sums usually depend on the $\varepsilon$ parameter. However, in most cases one will fail to solve the arising recurrences within this class. In particular, the definite sums inside of our method as outlined in Subsection~\ref{Subsec:DefiniteSums} cannot be expressed in our \rpisiE-extensions.

\noindent In the following we use the fact that the Laurent expansion of the Feynman integrals (and the underlying summation problems) around $\varepsilon =0$ to a finite order is of primary interest to the physics community. Consider~\eqref{Equ:HSum}; for simplicity, we will assume that $h_i=1$ for $1\leq i<m$ and we set $h=h_m$.
 Then a more flexible tactic is to focus on the Laurent expansion of
\begin{equation}\label{Equ:hExpand}
h(n,k_1 ,\ldots , k_m)=f_{l}(n,k_1 ,\ldots , k_m)\varepsilon^l+\dots+f_{r}(n,k_1 ,\ldots , k_m)\varepsilon^r+O(\varepsilon^{r+1})
\end{equation}
w.r.t.\ $\varepsilon$ up to a certain order $r$ with $r\geq l$; in 3-loop calculations one expects $l=-3$. If the sums in~\eqref{Equ:HSum} are finite, one obtains the first coefficients
\begin{equation}\label{Equ:SumFreeOfEp}
F_i(n)=\sum_{k_1=\alpha_1}^{L_1(n)} \cdots \sum_{k_m=\alpha_m}^{L_m(n,k_1,\ldots, k_{m-1})} f_i(n,k_1 ,\ldots , k_m)
\end{equation}
of the desired $\varepsilon$-expansion
$$S(n)=F_l(n)\varepsilon^l+\dots +F_r(n)\varepsilon^r+O(\varepsilon^{r+1}).$$
If also infinite sums are involved, extra care has to be taken into account.
As it turns out the $f_i$ themselves can be again written in terms of hypergeometric products together with harmonic numbers and cyclotomic harmonic sums~\cite{ABS:11}. Hence one option is to apply our summation methods to~\eqref{Equ:SumFreeOfEp} which is free of $\varepsilon$. Then in basically all our calculations the arising definite sums turn out to be solvable within our difference ring approach. However, the coefficients $f_i$ in~\eqref{Equ:hExpand} and thus the summands in~\eqref{Equ:SumFreeOfEp} get more and more involved (in particular they depend more and more on the harmonic sums) which blow up the calculations.

More successfully one can apply our new algorithms in combination with the following clever $\varepsilon$-expansion
technique~\cite{Ablinger:2012ph} to our simple example~\eqref{Equ:SimpleExpand}.
As an illustration of the refined approach for a particle physics sum consider the following
\begin{equation}\label{Equ:SimpleExpand}
S(n) = \sum_{k=0}^{n-2}\frac{n-k-1}{1+k}\sum_{j=0}^{n-k-2}\frac{  (-1)^{j} (k+j)!\Pochhammer{1-\frac{\varepsilon}{2}}{k}\Pochhammer{2-\frac{\varepsilon}{2}}{j}}{\Pochhammer{3-\varepsilon}{k+j}\Pochhammer{3+\frac{\varepsilon}{2}}{k+j}}\binom{n-k-2}{j}.
\end{equation}
We start to compute a refined holonomic system for the inner sum denoted by $b_{n,k}$:
\begin{equation}\label{Equ:InnerSystemEp}
\begin{split}
0=&-(1+k) (2
-\varepsilon
+2 k
) b_{n,k}
+\big(
        14
        +\varepsilon
        -\varepsilon^2
        +14 k
        +\varepsilon k\\
        &\quad\quad+4 k^2
        -\varepsilon n
        -2 k n
\big) b_{n,k+1}
-2 (2
+\varepsilon
+k
) (3
+k
-n
) b_{n,k+2},
\\
0=&\big(
        2
        -\varepsilon^2
        +2 k
        +\varepsilon k
        +2 k^2
        +2 n
        -\varepsilon n
        -2 k n
        +2 n^2
\big) b_{n,k}\\
-&2 (1
+\varepsilon
+k
) (2
+k
-n
) b_{n,k+1}
+(-1
+\varepsilon
-n
) (2
+\varepsilon
+2 n
) b_{n+1,k}.
\end{split}
\end{equation}
\normalsize
The complete double sum can be written as,
$S(n) = \sum_{k=0}^{n-2} \frac{n-k-1}{1+k} b_{n,k}.$
Using~\texttt{Sigma} once more the sequence obeys a recurrence only valid for the upper bound $n-4$. For this adjusted sum $S'(n)$ we get
\begin{multline}\label{Equ:SRecExpanded}
-2 (1+n)^2 (2+n) (2
+\varepsilon
+2 n
) S'(n)
\\
-(2+n) (2
+\varepsilon
+2 n
)  \big(
        -8
        +2 \varepsilon
        +\varepsilon^2
        -10 n
        +\varepsilon n
        -4 n^2
\big)S'(1+n)\\
+(1+n) (-2
+\varepsilon
-n
) (2
+\varepsilon
+2 n
) (4
+\varepsilon
+2 n
)S'(2+n)=r(\varepsilon,n)
\end{multline}
where $r(\varepsilon,n)$ depends on $b_{n,0},b_{n,1}$ and $b_{n,n-4},b_{n,n-3}$. Rewriting these definite sums, that depend on $\varepsilon$, to an expression  in terms of indefinite nested sums is not possible. Therefore, we compute the $\varepsilon$-expansion of the arising sums (e.g., by expanding the summands and applying the summation quantifiers to the coefficients of their expansion as proposed above).
Solving these telescoping points, i.e., computing the first coefficients of their $\varepsilon$-expansion gives
\begin{multline}\label{Equ:rExpanded}
r(\varepsilon,n)=
\tfrac{16 \big(
        336+48 n-248 n^2-70 n^3+186 n^4-121 n^5+138 n^6-81 n^7+26 n^8+7 n^9-6 n^{10}+n^{11}\big)}{(-2+n)^3 (-1+n)^3 n^2 (1+n)} \\[-0.1cm]
+\varepsilon \big[
        -\tfrac{4\big(
                -4032+7104 n+\ldots -7 n^{15}+n^{16}\big)}{(-2+n)^4 (-1+n)^4 n^3 (1+n)^2 (3+n)}
        -\tfrac{16 (-1+n) S_1({n})}{3+n}
\big]
\dots
\end{multline}
Finally, given the first initial values $F_i(j)$ with $i=0,1$ and $j=2,3$ in $S(2)=F_0(2)+F_1(2)\varepsilon+\dots$ and $S(3)=F_0(3)+F_1(3)\varepsilon+\dots$ one can activate \texttt{Sigma}'s $\varepsilon$-expansion solver~\cite{BKSF:12} to~\eqref{Equ:SRecExpanded} and obtains the coefficient $F_0(n)$ and $F_1(n)$ of $S'(n)=F_0(n)+F_1(n)\varepsilon+O(\varepsilon^2)$. Taking care of the extra points $k=n-2,n-3$
one finally obtains the expansion of the input sum $S(n)$. With the implementation \texttt{RhoSum} a complete automation is possible with the function call
\begin{mma}
\In FindSum[ \tfrac{(-1)^j(-1
-k
+n
) \binom{-2
-k
+n
}{j} (j
+k
)! \big(
        1-\frac{\varepsilon}{2}\big)_k \big(
        2-\frac{\varepsilon}{2}\big)_j}{(1+k)(3-\varepsilon)_{j
+k
} \big(
        3+\frac{\varepsilon}{2}\big)_{j
+k
}}
,\{\{j,0,n-k-2\},\{k,0,n-2\}\},\newline
\hspace*{6.cm}\{n\},\{3\},\{\infty\},ExpandIn\to \lbrace \varepsilon ,0 ,1 \rbrace ]\vspace*{-0.3cm}\\
\Out \big\{-4 S_1({n})
+8 n S_3({n})
-4 n S_{2,1}({n})
,\quad-\frac{(5+n)}{1+n}S_1({n})
-S_1({n})^2
+3 S_2({n})
+2 n S_2({n})^2\newline
-2 n S_3({n})
+6 n S_4({n})
+n S_{2,1}({n})-4 n S_{3,1}({n})
-2 n S_{2,1,1}({n})
\big\}\\
\end{mma}

\noindent Instead of calculating the expansions of the sums in~\eqref{Equ:rExpanded} in the old-fashioned way (i.e., by expanding the summands and applying the summation quantifiers to the coefficients of their expansion), one can apply recursively our proposed technology to obtain the expansion~\eqref{Equ:rExpanded}.

More generally, following the strategy in Section~\ref{Sec:MultiSumApproach}, we will calculate stepwise from inside to outside a refined holonomic system
given by the recurrences~\eqref{Equ:PureRec} and~\eqref{Equ:HookRec} but in each step we will expand the inhomogeneous parts in an $\varepsilon$-expansion whose coefficients can be expressed in an \rpisiE-extension. E.g., the inhomogeneous parts of the recurrences in~\eqref{Equ:InnerSystemEp} are $0$ and the $\varepsilon$-expansion is trivial.
Further, the recurrence~\eqref{Equ:SRecExpanded} with~\eqref{Equ:rExpanded} can be a component of a refined holonomic system that might be used to tackle another sum which is on top.

More practically,~\texttt{RhoSum} has been applied to solve many such sums being with many more summation quantifiers that
originate in particle physics. Here we would like to mention the calculation of the massive 3-loop contributions
to the heavy quark effects for the structure functions in deep-inelastic scattering \cite{Blumlein:2012bf}.
This has already contributed to a rich
literature~\cite{Ablinger:2010ty,Behring:2013dga,Behring:2014eya,Ablinger:2014uka,Ablinger:2014vwa,
Physics1,Physics2,Physics3,Behring:2015roa,Behring:2016hpa,Ablinger:2017tan,Ablinger:2017err}
which has made new physical insight possible~\cite{Physics4}. When this project is completed,
the strong coupling constant $\alpha_s(M_Z^2)$ and the charm quark mass $m_c$ can be measured from the deep-inelastic
world data with unprecedented accuracy and a significant improvement of the gluon distribution function of the
nucleon can be achieved. This has important consequences for all precision measurements at the LHC (Large Hadron Collider, CERN), because
these quantities determine the QCD corrections to the corresponding production cross sections.

\section{\label{Summ}Summary}
A new method of summation known as the refined holonomic approach has been introduced that extends significantly
the ideas worked out in~\cite{Schneider:05a}. Its main features are the ability to work with inhomogeneous
recurrences and to balance the difficulty of a telescoping problem with the number of ring extensions made to
find a solution. These techniques have proved useful in particle physics where inhomogeneous recurrences
are essential to modern computation. A future development may consist of the application of the algorithm
to summation problems in which further real parameters, beyond the dimensional parameter $\varepsilon$, are
present. This is of high relevance for multi-leg scattering processes at high energy colliders like the LHC and
a planned future $e^+e^-$ collider. These parameters
may be the masses and/or the virtualities of the external legs of the corresponding Feynman diagrams. \texttt{RhoSum} is
expected to handle problems of this kind more efficiently than other implementations.
The algorithm has been implemented in~\texttt{Mathematica} to employ the technique in practical situations and problems from particle physics involving large numbers of sums have been successfully solved.\\
The earlier approach~\cite{Schneider:05a} served as the central tool to provide the first computer assisted proof~\cite{APS:05} of Stembridge's celebrated TSPP theorem~\cite{STEMBRIDGE1995227} in the context of plane partitions. In that time the computation steps have been carried out manually. First experiments show that our new package \texttt{RhoSum} in interaction with \texttt{Sigma} can support the user heavily: many steps can be carried out now mechanically and critical special cases are discovered automatically. In a nutshell, our tools can guide the user to a big extend through these complicated and subtle calculations. It is expected that this machinery will contribute further in difficult calculation combing from particle physics but will will also assist in new challenging problems in the context of combinatorics.

\medskip

\noindent{\bf Acknowledgment.}
This work was supported in part by the guest programme of Kolleg Mathematik-Physik Berlin (KMPB), the  European Commission through contract PITN-GA-2012-316704 ({HIGGSTOOLS}), and by the Austrian Science Fund (FWF) grant SFB F50 (F5009-N15).


\end{document}